\documentclass{aa} 
\usepackage[breaklinks=true, colorlinks=true, linkcolor=blue, citecolor=blue]{hyperref}
\usepackage{graphicx}
\DeclareGraphicsExtensions{.pdf,.png,.jpg,.ps,.eps}
\usepackage{xcolor}
\usepackage{ltablex} 
\usepackage{placeins} 
\usepackage{lscape}
\usepackage{color}
\usepackage[varg]{txfonts}

\begin{document}

   \title{Oxygen and silicon abundances in Cygnus OB2}

\subtitle{Chemical homogeneity in a sample of OB slow rotators}

   \author{S. R. Berlanas
          \inst{1,2}
         \and
          A. Herrero\inst{1,2}
          \and
          F. Comerón\inst{3}
          \and
          S. Simón-Díaz\inst{1,2}
          \and
          M. Cerviño\inst{4}
          \and
          A. Pasquali\inst{5}         
         }

   \institute{Instituto de Astrofísica de Canarias, 38200 La Laguna, Tenerife, Spain\   			              
         \and Departamento de Astrofísica, Universidad de La Laguna, 38205 La Laguna, Tenerife, Spain\	            
         \and ESO, Karl-Schwarzschild-Strasse 2, 85748 Garching bei München, Germany\ 
         \and Centro de Astrobiolog\'ia (CSIC/INTA), 28850 Torrej\'on de Ardoz, Madrid, Spain\
         \and Astronomisches Rechen-Institut, Zentrum für Astronomie der Universität Heidelberg, Mönchhofstr 12–14, 69120 Heidelberg, 
         Germany 
             }

   \date{Received month day, year; accepted month day, year}

   \abstract
   { Cygnus OB2 is a rich OB association in the Galaxy located at only $\sim$1.4 kpc from us which has experienced intense star formation in the last 20-25 Myr. Its stellar population shows a correlation between age and Galactic longitude. Exploring the chemical composition of its stellar content we will be able to check the degree of homogeneity of the natal molecular cloud and possible effects of self-enrichment processes.}
   {Our aim is to determine silicon and oxygen abundances for a sample of eight early-type slow rotators (with rotational velocities below 80 km~s$^{-1}$) in Cygnus OB2 in order to check possible inhomogeneities across the whole association and whether there exists a correlation of chemical composition with Galactic longitude.}
   {We have performed a spectroscopic analysis of a sample of late O and early B stars with low rotational velocity in Cygnus OB2, which have been chosen so as to cover the whole association area. We have carried out an analysis based on equivalent widths of metal lines, the wings of the H Balmer lines and FASTWIND stellar atmosphere models to determine their stellar fundamental parameters (effective temperature and surface gravity) as well as the silicon and oxygen surface abundances.}
   {We derive a rather homogeneous distribution of silicon and oxygen abundances across the region, with average values of 12 + log(Si/H) =  7.53 $\pm$ 0.08 dex and 12 + log(O/H) = 8.65 $\pm$ 0.12 dex.}
   {We find a homogeneous chemical composition in Cygnus OB2 with no clear evidence for significant chemical self-enrichment, despite indications of strong stellar winds and possible supernovae during the history of the region. Comparison with different scenarios of chemical enrichment by stellar winds and supernovae point to star forming efficiencies not significantly above 10$\%$. The degree of homogeneity that we find is consistent with the observed Milky Way oxygen gradient based on \ion{H}{II} regions. We also find that the oxygen scatter within Cygnus OB2 is at least of the same order than among \ion{H}{II} regions at similar Galactocentric distance.}


   \maketitle
%

\section{Introduction}

The chemical composition of the natal molecular cloud is one of the fundamental parameters in the process of massive star formation, as it affects the opacity of the infalling matter, the chemical processes that take place during star formation or the upper mass limit, to mention some \citep[e.g.][]{zy07, tan14}. The previous enrichment history of the Interstellar Medium (ISM) through stellar evolution and processes like stellar winds, metal-rich yields from supernovae (SNe) or infall of metal-poor gas, directly affects the chemical composition of the natal molecular cloud \citep[e.g.][]{langer12, prantzos18, barrera18}. 

Young open clusters and OB associations have been proposed as places to explore the effects of self-enrichment by stellar winds and SNe \citep[e.g.][]{spina17, palla11, biazzo11a, ssimon10, cunha94, cunha92} as many of them have a star formation history sufficiently long to have experienced the explosion of their oldest massive members as SNe, contaminating the environment through processes driven by the pollution of the interstellar medium by subsequent generations of massive stars. We may thus expect that subsequent stellar generations show chemical differences. Inversely, the presence or absence of these differences allow us to put constrains on the processes involved in the chemical pollution of the proto-stellar cloud. For example, predictions about the oxygen yields from core-collapse supernovae (CCSN) greatly differ depending on the  upper mass limit to become CCSN or on the physics of massive stars included in the models \citep[see e.g.][and references therein]{suzuki18, prantzos18}.
  Unfortunately, the stellar content of many OB associations is difficult to study in detail due to the combined effects of distance and extinction.
  
A study in Orion \citep{ssimon10} resulted in a chemically homogeneous composition, contrary to previous works claiming evidence of self-enrichment \citep{cunha94}. The homogeneity has later been confirmed by other independent analysis strategies \citep[see][]{nieva11,cunha12}. Moreover, this result was not only found in early-type stars, but also in late-type stars \citep[e.g.][]{biazzo11a,biazzo11b}. But Orion is a modest region in terms of star formation. 

Cygnus OB2, because of its close distance of $\sim$1.4 kpc \citep{rygl12} and relatively intense star formation, provides an excellent target to study possible chemical composition inhomogeneities through self-enrichment. Massive star formation seems to have peaked around 3 Myr ago \citep{com12,wright15, berlanas17} and the distribution of stellar ages extends beyond 20 Myr \citep{com16}. Although there are no known supernova remnants clearly associated with Cygnus OB2, the current existence of a large number of B-type giants  \citep{berlanas17} and seven red supergiants in the region \citep{com16} indicates an evolved population whose most massive members have already exploded as SNe. The three Wolf-Rayet stars known in the core of the region  and the large number of OB stars \citep{wright15}, represent intense stellar wind sources susceptible to produce chemical enrichment across the association.
Strong signatures from the decay of  $^{26}$Al in Cygnus OB2 \citep{knoedl02}, other high-energy detections at $\gamma$- and X-rays \citep{martin9, martin10, bednarek03, butt06} or the presence of the young pulsar J2032+4130 in the surroundings \citep{abdo09, camilo09}, could have been caused chemical enrichment produced by stellar winds and/or SNe.

In virtue of its rich content in massive stars and an extended star formation history sufficiently long to have included the explosion of its oldest massive members, Cygnus OB2 represents a suitable region to explore possible abundance inhomogeneities. The observed spatial age gradient across its extent \citep{com12} might be associated with a chemical composition gradient produced by pollution of the ISM by successive generations of massive stars.

In Sect.~\ref{sect2} we describe the target selection and the observing strategy. In Sect.~\ref{sect3} we present the whole spectroscopic analysis, including the derived EWs, stellar parameters and abundances of a sample of eight early-type slow rotators of Cygnus OB2. In Sect.~\ref{sect4} we  discuss our results and their implications. Finally, in Sect.~\ref{sect5}, we summarize the main conclusions of this work.

\section{Target selection, observations and membership}\label{sect2}

Late O-type and early B-type stars, with their rich spectrum of silicon and oxygen lines are ideal targets to test possible chemical contamination. These stars should cover the whole association area (1 deg. radius centered on Galactic coordinates $l = $79.8$^{\circ}$ and $b = $+0.8$^{\circ}$).
In addition, to reach an optimal accuracy in the abundance analysis, we need to analyze stars with effective temperature in the range 18000\,--\,34000 K  or, equivalently, spectral types in the range B2\,--\,O9.5 (where silicon ionization balance and \ion{O}{II} lines are strong enough , see \cite{ssimon10}) and low projected rotational velocities (in what follows, we call slow rotators those stars with projected rotational velocities below 80 km~s$^{-1}$) at high resolution and signal-to-noise ratio (S/N). The fraction of early-B dwarfs and giants rotating at these velocities depends on the environment and evolutionary stage of the cluster considered, but from the results in the Milky Way \citep{dufton06, wolff07, huang06} and 30 Dor \citep{dufton13} we may expect it to be 20 - 30$\%$, which makes mandatory to previously observe a large sample of OB stars to identify early-type slow rotators.

In this section we describe the observations and various steps we have performed in order to build an optimal list of stars for the purposes of this paper. Briefly we started with intermediate resolution observations of a large list of OB stars in Cygnus OB2 which helped us  identifying the late-O and early-B slow rotators. Then we proceeded to gather new high-resolution observations of this optimized sample. In the last step, we benefit from data about parallaxes and proper motions from the second Gaia data release \citep[Gaia DR2, ][]{gaia18} to asses the membership of the final stellar sample. 

  \begin{table}[t!]
	\centering
	\caption{Coordinates (J2000.0) of the center of the fields observed in July 2014 using WYFFOS at WHT.}
	\label{table1}
		\begin{tabular}{ccc}
		\hline 
		\hline 
   		 \small{Field}& \small{RA (hhmmss)}& \small{Dec ($^{\circ}$ $^\prime$ $^{\prime\prime}$)} \\
   		\hline 
      	\small{Cygnus OB2 - field1} & \small{20 33 04} & \small{41 27 00}\\
		\small{Cygnus OB2 - field2} & \small{20 31 03} & \small{40 36 20} \\
 		\small{Cygnus OB2 - field3} & \small{20 30 03} & \small{41 01 20} \\
 		\small{Cygnus OB2 - field4} & \small{20 34 50} & \small{40 33 20} \\
      	\hline
		\end{tabular}	
\end{table}

 \begin{figure}[t!]
\centering
\includegraphics[width=8.8cm]{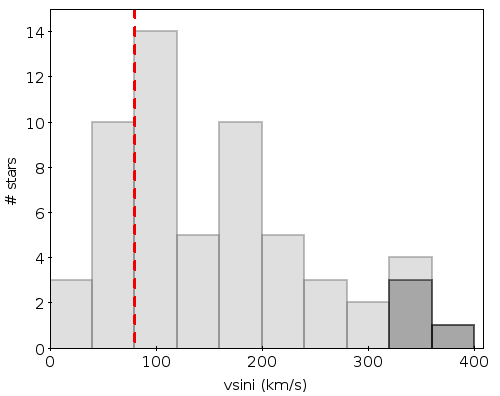}
\caption{ Rotational velocity distribution of a sample of 57 OB-type stars in Cygnus OB2. Dark gray color indicates possible binary candidates. The red dashed line indicates $v$ sin $i =$ 80 km~s$^{-1}$.}
\label{fig2}
\end{figure}

\subsection{Intermediate resolution spectroscopy and early-type slow rotators selection}\label{sect21}

\begin{figure*}[t!]
\centering
\includegraphics[width=13.0cm]{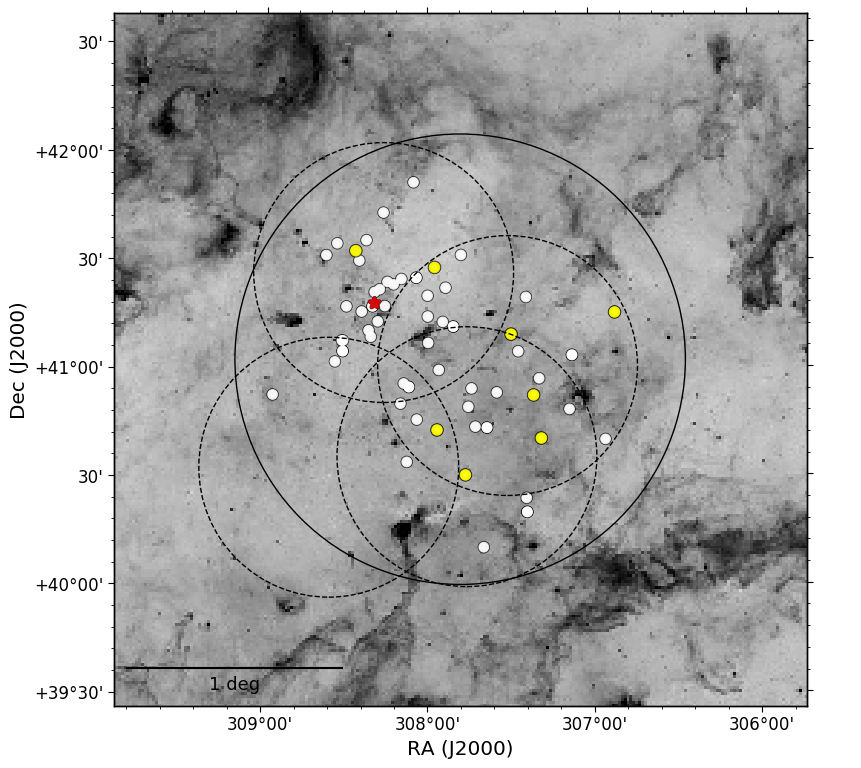}
\caption{ Inverse Spitzer 8 $\mu$m image showing the location of the 57 stars (dots) observed in Cygnus OB2, and the selected OB slow rotators (yellow dots) suitable for the abundance analysis. The solid line circle indicates the 1 deg. Cygnus OB2 area adopted in this work, and the four dashed line circles the observed AF2/WYFFOS fields. For reference, the red star indicates the location of the Cyg OB2 $\#8$ Trapezium-like system. }
\label{fig3}
\end{figure*}

 \begin{table*}[t!]
	\centering
	\caption{Sample of the selected eight early-type slow rotators observed with ISIS at the WHT. Columns give their names, spectral types (from this and other works), B magnitudes, coordinates (J2000.0 epoch) and derived rotational velocities. The spectral signal to noise (per pixel) ratio and  resolving power (per element) are also indicated. B magnitudes are taken from \cite{com12}. (b) and (r) refer to the blue and red optical range spectra respectively.}
	\label{table2}
		\begin{tabular}{lcccccccccc}
		\hline   
		\hline\\[-1.8ex]
    	\small{Object} & \small{SpT (this work)}&  \small{SpT (other works)}&\small{$B$ (mag)}&\small{Coordinates ($^{\circ}$)} &\small{$v$ sin $i$ (km~s$^{-1}$)}& \small{S/N(b)}& \small{S/N(r)}& \small{R(b)}& \small{R(r)}\\
    \hline	\\[-1.5ex]
\small{J20334610+4133010}  &\small{O9.7 Iab} &\small{O9.7 Iab$^{1}$}& \small{11.2}& \small{308.44 +41.55}& \small{55}& \small{180}& \small{230}& \small{13600}& \small{9300}\\    
\small{J20295701+4109538}  &  \small{O9.7 III} &\small{B0V$^{2}$}& \small{13.0}& \small{307.49 +41.16}& \small{70}& \small{110}& \small{160}& \small{13600}& \small{9300}\\ 
\small{J20314965+4128265}  &\small{O9.7 IV}&\small{O9III$^{3}$} & \small{12.6}& \small{307.96 +41.47}& \small{30}& \small{180}& \small{230}& \small{13600}& \small{9300}\\       
\small{J20272428+4115458}  &\small{B0 IV} &\small{O9.5V$^{4}$}& \small{12.6}& \small{306.85 +41.26}& \small{20}& \small{110}& \small{220}& \small{13600}& \small{9300}\\
\small{J20292449+4052599}  & \small{B0.2 IV} &\small{B0.2 IV$^{4}$}& \small{13.8}& \small{307.35 +40.88}& \small{75}& \small{100}& \small{130}& \small{7500}& \small{9300}\\
\small{J20314605+4043246}  & \small{B0.5 IV} &\small{B0.5 IV$^{2}$}& \small{13.8}& \small{307.94 +40.72}& \small{60}& \small{100}& \small{150}& \small{7500} & \small{9300}\\
\small{BD+40 4210}  &  \small{B2 Ia} &\small{B1III:e$^{4}$}& \small{12.2}& \small{307.77 +40.52}& \small{30}& \small{130}& \small{150}& \small{13600}& \small{9300} \\
\small{BD+40 4193}  &  \small{B2 V}&\small{B1V$^{5}$} & \small{10.4}& \small{307.31 +40.68} & \small{70}& \small{150}& \small{250}& \small{13600}& \small{9300}\\ 
\hline
		\end{tabular}
		\tablefoot{Spectral type source: 1) \cite{maiz16}, 2) \cite{hanson03}, 3) \cite{kiminki07}, 4) \cite{com12}, 5) \cite{berlanas17}.}
\end{table*}

In July 2014 we obtained intermediate-resolution spectroscopy of 51 stars in Cygnus OB2 classified as late-O and early-B in the large list from \cite{com12} using the multi-object AF2/WYFFOS spectrograph at the William Herschel Telescope (WHT). We selected four fields in the region (see Table~\ref{table1}) in order to maximize the number of suitable candidates. 
Simulations based on a B1V spectrum broadened by different values of $v$ sin $i$ and different levels of noise indicated that the best trade-off between the required exposure time and enough $v$ sin $i$ accuracy for our target selection can be achieved with a combination of a resolving power of 5000 and S/N $\geq$ 50. We thus have chosen the H2400B grating, which provides the required resolution and wavelength coverage, setting the central wavelength to 4500 $\AA$. 
The spectral region contains the \ion{Si}{III} lines at  $\lambda$4552,  $\lambda$4568 and  $\lambda$4575 $\AA$, which are broadened mainly by rotation and are most prominent around the B1V type stars.
Finally, in order to improve the statistics of our sample, we added six stars from previous unpublished observations in Cygnus OB2 carried out by A. Herrero in August 2012 at the WHT using the ISIS spectrograph with the R1200B and R1200R gratings (covering all the lines of interest and providing high quality spectra good enough for our purposes).

We determined the projected rotational velocities for  the whole stellar sample of 57 stars using {\small{\texttt{iacob\_broad}}}, an user-friendly tool for OB-type spectra which is based on a combined Fourier transform plus a goodness-of-fit methodology \citep{ssimon07, ssimon14}. We have mainly based the analysis on the \ion{Si}{III} $\lambda$4552 line or the \ion{N}{III} $\lambda$4510 line in the cases in which the \ion{Si}{III} line was weak. When both lines were weak, we have used the \ion{He}{I} $\lambda$ 4387,4471 lines. Typical uncertainties are in the order of 10-20\% \citep{ssimon14}.
 We found a fraction of OB slow rotators consistent with the expectations, reaching a $\sim$23$\%$ of the targets at $v$ sin $i \lesssim$ 80 km~s$^{-1}$ (see Fig.~\ref{fig2}). For four targets, represented with dark gray color, we suggest possible spectroscopic binarity (two of them are reported as binary candidates by \cite{berlanas17}). We can only suggest their possible binary nature due to their noisy spectra. 
 
 The whole list of 57 stars (note that some of them are too hot for the abundance analysis) with the derived rotational velocities is given in Appendix~\ref{app_vsini}.
From this sample, and taking into account the required conditions mentioned above, six stars were suitable for our abundance analysis.

\begin{table*}[t!]
	\centering
	\caption{Gaia sources, parallaxes and proper motions from the second Gaia data release. Parallax errors do not contain the systematic uncertainty discussed by \cite{arenou18,lindegren18,stassun18}.}
	\label{table_distances}
		\begin{tabular}{lcccc}
		\hline   
		\hline\\[-1.8ex]
    	\small{Target}& \small{Gaia name} & \small{parallax (mas)}& \small{pm$_{\rm RA}$ (mas yr$^{-1}$)}& \small{pm$_{\rm Dec}$ (mas yr$^{-1}$)}  \\
 \hline \\[-1.5ex]

\small{J20314965+4128265}& \small{2067847502568383744} & \small{ 0.595$\pm$0.032 } &\small{-3.09$\pm$0.05 } &\small{-4.18$\pm$0.04 } \\ 
\small{J20295701+4109538}& \small{2067816681882846464} & \small{0.587$\pm$0.024} &\small{-3.15$\pm$0.04 } & \small{-4.58$\pm$0.04} \\ 	
\small{J20272428+4115458}& \small{2067642233192106624} & \small{0.349$\pm$0.027} &\small{-2.96$\pm$0.04} &\small{-3.67$\pm$0.05}  \\
\small{J20334610+4133010}& \small{2067928660269306496}& \small{0.590$\pm$0.449} &\small{1.21$\pm$0.85} & \small{-1.46$\pm$0.77} \\
\small{J20314605+4043246}& \small{2067742877164542080} & \small{0.377$\pm$0.256} &\small{-1.26$\pm$0.48} & \small{-4.38$\pm$0.42}  \\  
\small{J20292449+4052599}& \small{2067429168454304896} & \small{0.585$\pm$0.027} &\small{-2.24$\pm$0.04} & \small{-3.71$\pm$0.05} \\   
\small{BD+40 4193}& \small{2067425461902141056}& \small{0.750$\pm$0.035} & \small{-0.39$\pm$0.05} &\small{-2.38$\pm$0.06}  \\ 
\small{BD+40 4210}& \small{2067734012352626816}& \small{0.653$\pm$0.057} &\small{-3.17$\pm$0.09} &\small{-4.61$\pm$0.11}  \\
\hline
		\end{tabular}
\end{table*}

\subsection{High resolution and S/N spectroscopy}\label{sect22}

 In July 2015 we obtained high resolution, high S/N spectra of the six selected stars. We obtained ISIS spectra at the WHT using both, the blue and red arms. We chose the H2400B grating for the blue arm, which provides the required resolving power (R$\sim$13600 at 4500$\AA$).
The spectral coverage of the H2400B grating requires the observation of each star in two spectral ranges centered on 4200 and 4600$\AA$, in order to cover all the lines of interest. But due to bad weather conditions and in order to get high S/N spectra, the H2400B grating was replaced by the R1200B one for the faintest targets (J20292449+4052599 and J20314605+4043246), which provides a lower but still sufficient resolving power (R$\sim$7500 at 4500$\AA$ by reducing the slit width to 0.7 arcsec) and covers all lines of interest. As the dichroic allows to simultaneously observe the red arm, we used the R1200R grating (R$\sim$9300 at 6500$\AA$) centered at 5700 and 6500 Å for all the stars, covering the regions of \ion{O}{III} $\lambda$5592, \ion{C}{IV} $\lambda$5811-14 $\AA$ and H$\alpha$. 
Additionally we have included two more stars, J20314965+4128265 and J20334610+4133010, both OB slow rotators that meet all the requirements and were observed by A. Herrero in July 2003 using also ISIS at the WHT with the same configuration (H2400B and R1200R gratings).
 Figure~\ref{fig3} shows the location of the whole sample of 57 OB stars in the region, and the eight stars finally chosen to carry out the abundance analysis.

The spectra were reduced using IRAF\footnote{IRAF is distributed by the National Optical Astronomy Observatories, which are operated by the Association of Universities for Research in Astronomy, Inc., under cooperative agreement with the National Science Foundation.} with standard routines for bias and flat-field subtraction, and also for the wavelength calibration.  A summary of their basic data is presented in Table~\ref{table2}.

\subsection{Gaia-DR2 data}\label{sect42}
 
We used the recent data (parallaxes and proper motions) from Gaia DR2 to check whether all the stars of our sample belong to the same region and present no peculiarities in the data that could indicate a different origin or evolutionary history.

All but one of the stars in our sample  with a parallax uncertainty below 0.1 mas (see Table~\ref{table_distances}) have values consistent with the known Cygnus OB2 distance of 1.4 - 1.7 kpc \citep{mt91, hanson03,rygl12}. There is one star (J20272428+4115458) which seems to be at further distance (about 1.8 sigma from the mean parallax value of our sample). 
However, there is a systematic (but variable) offset up to 0.1 mas that could affect all the reported Gaia DR2 parallaxes \citep{arenou18,lindegren18,stassun18}. Nevertheless, the oxygen abundance obtained for this star is consistent with the mean value obtained for the sample. 
Other two stars have large parallax uncertainties, but still consistent with the average value.

Focusing on proper motions, they are all consistent with membership in Cygnus OB2. Our stellar sample with proper motion uncertainties below 0.1 mas yr$^{-1}$ have  mean values of pm$_{\rm RA}$=  -2.5 $\pm$ 0.05 mas yr$^{-1}$ and  pm$_{\rm Dec}$= -3.85 $\pm$ 0.06 mas yr$^{-1}$.

We therefore conclude that our sample belongs to the normal population of Cygnus OB2 and does not include any runaway star.

  \begin{table}[b!]
	\centering
	\caption{Stellar parameters derived for the sample of stars by using the \ion{Si}{III-IV} and/or \ion{Si}{II-III} ionization equilibrium along with the H Balmer lines. $T_{\rm eff}$ and log $g$ uncertainties are, in all cases, 500 K and 0.1 dex respectively. An asterisk indicates that $T_{\rm eff}$ has been obtained using the \ion{He}{I-II} ionization balance. }
	\label{table3a}
		\begin{tabular}{lccc}
		\hline   
		\hline\\[-1.8ex]
    	\small{Target} & \small{SpT}& \small{$T_{\rm eff}$ (K)}& \small{log $g$ (dex)} \\
 \hline\\[-1.5ex] 

\small{J20314965+4128265} & \small{O9.7 IV} &\small{33000}& \small{4.0}\\ 
\small{J20295701+4109538} & \small{O9.7 III} & \small{32000}& \small{3.6}\\ 	
\small{J20272428+4115458} & \small{B0 IV}& \small{30000}& \small{3.9}\\
\small{J20334610+4133010} & \small{O9.7 Iab}&  \small{30000}& \small{3.2}\\
\small{J20314605+4043246} & \small{B0.5 IV}&  \small{30000}& \small{4.0}\\  
\small{J20292449+4052599} & \small{B0.2 IV}&  \small{29000*}& \small{3.7}\\   
\small{BD+40 4193} & \small{B2 V}&   \small{19000}& \small{3.8}\\ 
\small{BD+40 4210} & \small{B2 Ia}&   \small{18300}& \small{2.2}\\
\hline
		\end{tabular}
\end{table}

\section{Spectroscopic analysis}\label{sect3}

We have followed the same methodology as \cite{ssimon10}, where the details of the FASTWIND grid used (that includes \ion{H}{I}, \ion{He}{I-II}, \ion{Si}{II-III-IV} and \ion{O}{II}) and the considered atomic models are provided. In addition, we have extended the grid to lower gravities for the supergiants analysis.

 \subsection{Spectral classification}\label{sect30} 
 
We have reclassified the eight selected stars since the new spectra obtained in this work have, for many of them, a higher quality than in previous studies.
We have classified the O-type stars using the MGB tool \citep{maiz12}. It compares the observed spectra with a grid of O standards (in this work the GOSSS library, see \cite{maiz16}), allowing us to vary spectral type, luminosity class, velocity and resolution until obtaining a best match. For B-type stars we used templates gathered in the framework of the IACOBsweG spectroscopic survey \citep[][Negueruela et al. (in prep)]{ssimon15} to obtain the spectral types. 
The new classification of the selected stellar sample is presented in Table~\ref{table2} together with that from other works.

  \begin{figure}[t!]
\centering
\includegraphics[width=7.5cm,  height=6.5cm  ]{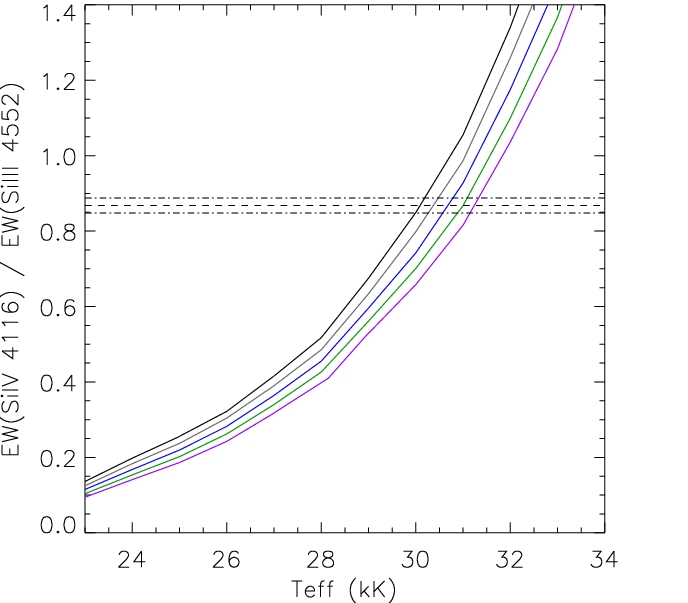} 
\includegraphics[width=4.3cm,  height=3.7cm]{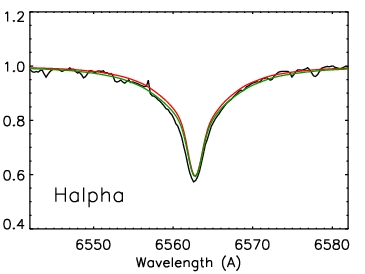}
\includegraphics[width=4.3cm,  height=3.7cm]{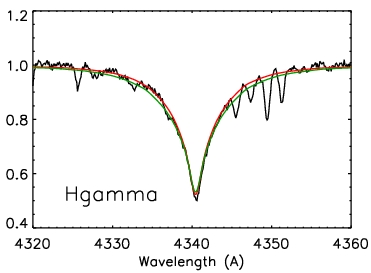}
\includegraphics[width=4.3cm,  height=3.7cm]{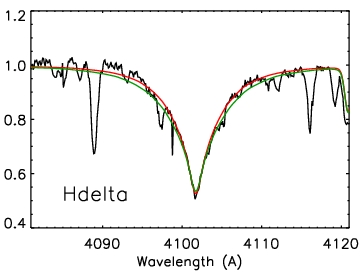}
\caption{ \textit{Top}: Example of the EW ratio of \ion{Si}{III-IV}  used as $T_{\rm eff}$ indicator with FASTWIND models for the star J20272428+4115458. Black, gray, blue, green and purple lines correspond to log $g$ values of 3.9, 4.0, 4.1, 4.2 and 4.3 dex respectively. The horizontal dashed line indicates the EW ratio of the star and its associated uncertainty is indicated by the dash-dotted lines. \textit{Bottom}: Example of the H Balmer lines fitting to determine log $g$ for the same star. Only two FASTWIND models are compared with the observed spectrum (black line) for clarity. In red, the $T_{\rm eff}$=30000 K, log $g$=3.9 dex model. In green, the $T_{\rm eff}$=31000 K, log $g$=4.2 dex model. Final fits can be seen in Figs.~\ref{sp_models1}-\ref{sp_models8}.}
\label{fig5a}
\end{figure}

	\subsection{Equivalent Widths}\label{sect31}
We have carried out an analysis based on equivalent widths (EW) of metal lines, similar to the method used in Orion by \cite{ssimon10}. We used our own IDL\footnote{Interactive Data Language
Proprietary software distributed by Research Systems, Inc. of Boulder, CO now a division of Kodak.} routines to identify metal lines in the observed spectra, and to measure the EWs and their uncertainties. 
For each line, a multi-Gaussian fit of the observed line profile is done in a spectral range  of $\lambda_o$ $\pm$ 2 max[$v$ sin $i$ $\lambda_o$/c, 0.5$\lambda_o$/R] around the laboratory wavelength of the line. The uncertainty is obtained by assuming the continuum at $\pm$1/SNR, and comparing the value obtained by means of the Gaussian fitting with the value derived by integrating the line profile. Measured EWs are given in Appendix~\ref{app}.

	\subsection{Stellar Parameters}\label{sect32}

    \begin{figure}[t!]
\centering
\includegraphics[width=8.0cm, height=9.3cm ]{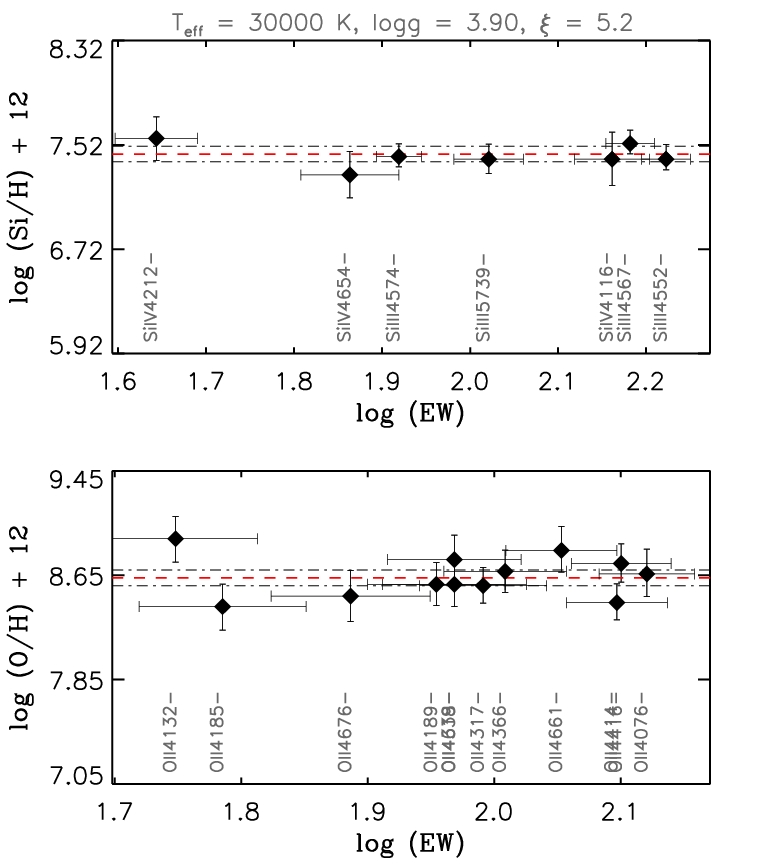}
\caption{ Example of the silicon (\textit{top}) and  oxygen (\textit{bottom}) abundance versus $EW$ diagrams for the star J20272428+4115458. All the observed \ion{Si}{III-IV} and \ion{O}{II} lines are indicated in the plots. The red dashed line indicates the weighed mean abundance (fit to the data), and the gray dash-dotted line the standard weighed deviation.  
}
\label{fig8}
\end{figure}

 Both parameters, effective temperature ($T_{\rm eff}$) and surface gravity (log $g$), were obtained by comparing the EW ratio of \ion{Si}{II-III} or \ion{Si}{III-IV} (depending on the effective temperature of the star) and the wings of the H Balmer lines with FASTWIND stellar atmosphere models \citep{santolaya97, puls05}. We first used the silicon line ratios to obtain the initial values of $T_{\rm eff}$ for different log $g$  values (e.g. Fig.~\ref{fig5a}, top). Then, the models obtained by setting each set of these parameters were compared with the wings of the observed H Balmer lines (e.g. Fig.~\ref{fig5a}, bottom) to iteratively obtain the final values for both parameters.

We have used  the ratios EW(\ion{Si}{IV} $\lambda$4116)/EW(\ion{Si}{III} $\lambda$4552) and EW(\ion{Si}{II} $\lambda$4130)/EW(\ion{Si}{III} $\lambda$4552) as $T_{\rm eff}$ indicators, depending on the temperature of the star,  because these are the strongest, unblended lines typically present in the stellar spectrum (at the spectral types of our targets  and in the range of $\sim$ 4000–5000 $\AA$). 
On the other hand, due to the good agreement between $T_{\rm eff}$ determined using \ion{Si}{III-IV} and the \ion{He}{I-II} ionization balance found by \cite{ssimon10},  the \ion{He}{I-II} ionization equilibrium has been used for the star J20292449+4052599 because only \ion{Si}{III} lines were suitable for the stellar parameter determination in the observed spectrum.

\begin{table*}[t!]
	\centering
	\caption{ Microturbulences ($\xi$) and final abundance values ($\epsilon_{x}$, in units of 12 + log(X/H)) derived for our stellar sample. Abundance uncertainties associated with microturbulence ($\Delta\epsilon(\xi)$), the standard weighted deviation in the line-by-line abundances ($\Delta\epsilon(\sigma)$), and those related to the stellar parameters ($\Delta\epsilon(T_{\rm eff},$log $g)$) are indicated separately.}
	\label{table3b}
		\begin{tabular}{lccccccccc}
		\hline   
		\hline\\[-1.8ex]
    	\small{Target}& \small{$\xi$} & \small{$\epsilon_{Si}$}& \small{$\Delta\epsilon_{Si}(\sigma)$}& \small{$\Delta\epsilon_{Si}(\xi)$}& \small{$\Delta\epsilon_{Si}(T_{\rm eff},log g)$} & \small{$\epsilon_{O}$}& \small{$\Delta\epsilon_{O}(\sigma)$}& \small{$\Delta\epsilon_{O}(\xi)$}& \small{$\Delta\epsilon_{O}(T_{\rm eff},log g)$}  \\
 \hline \\[-1.5ex]

\small{J20314965+4128265}& \small{3.0} & \small{7.67} &\small{0.06} &\small{0.08} &\small{0.07} & \small{8.75} & \small{0.07} &\small{0.02} &\small{0.11}\\ 
\small{J20295701+4109538}& \small{7.5} & \small{7.61} &\small{0.09} & \small{0.12} &\small{0.07} & \small{8.74} & \small{0.10} &\small{0.07} &\small{0.13}\\ 	
\small{J20272428+4115458}& \small{5.2} & \small{7.45} &\small{0.06} &\small{0.07} &\small{0.05} &  \small{8.63} & \small{0.06} &\small{0.12} &\small{0.07}\\
\small{J20334610+4133010}& \small{17.0}& \small{7.54} &\small{0.02} & \small{0.04} &\small{0.08} & \small{8.57} & \small{0.07} &\small{0.10} &\small{0.09}\\
\small{J20314605+4043246}& \small{5.6} & \small{7.54} &\small{0.09} & \small{0.12} &\small{0.05} & \small{8.82} & \small{0.10} &\small{0.12} &\small{0.02}\\  
\small{J20292449+4052599}& \small{6.8} & \small{7.47} &\small{0.07} & \small{0.10} &\small{0.12} & \small{8.51} & \small{0.15} &\small{0.02} &\small{0.05}\\   
\small{BD+40 4193}& \small{5.9}& \small{7.44} & \small{0.11} &\small{0.12} &\small{0.10} & \small{8.70} & \small{0.20} &\small{0.13} &\small{0.10}\\ 
\small{BD+40 4210}& \small{13.3}& \small{7.53} &\small{0.01} &\small{0.09} &\small{0.07} &  \small{8.48} & \small{0.06} &\small{0.07} &\small{0.10}\\
\hline
		\end{tabular}
\end{table*}

\begin{figure*}[t!]
\centering
\includegraphics[width=18.5cm]{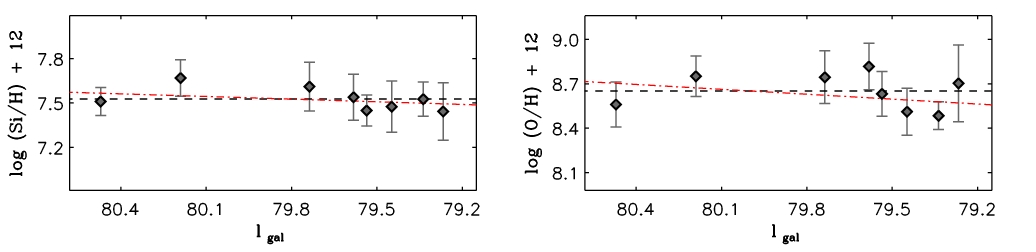}
\caption{ Derived silicon (\textit{left}) and oxygen (\textit{right}) abundance versus Galactic longitude in degrees. Dashed black lines indicate the mean abundance value and dash-dotted red lines the best fit to the data. Vertical lines indicate abundance errors, that include uncertainties related to the stellar parameters, microturbulence and the standard weighed deviation in the line-by-line abundances.}
\label{fig10} 
\end{figure*}

Derived $T_{\rm eff}$ uncertainties from $EWs$ of \ion{Si}{II-III-IV} lines are, for all our stars, lower than 500K. For the surface gravity we obtain an accuracy better than 0.1 dex. Taking into account the additional uncertainty sources (like continuum normalization) we follow the reasoning by \cite{sabin17} and adopt in all cases errors of $\pm$ 500 K and $\pm$ 0.1 dex for $T_{\rm eff}$  and log $g$, respectively. Derived values are shown in Table~\ref{table3a}.

	\subsection{Silicon and Oxygen abundances}\label{sect33}
	
	We applied the curve-of-growth method to derive silicon and oxygen abundances, considering a grid of HHeSiO FASTWIND models for a set of stellar parameters ($T_{\rm eff}$ and log $g$). For each star, the  abundance is obtained for different values of microturbulence and for each line considered. Then, we determined the final abundance value considering the microturbulence that, for all lines, gives a similar value (see Fig.~\ref{fig8}). More details can be found in e.g. \cite{kilian92} or \cite{ssimon10}.	
We have used all the lines available in our spectra, except those too faint or blended with other lines and that can not be properly separated by our IDL routines (see Sect.~\ref{sect31}) at the resolving power of the analyzed spectra. This is, for example, the case for the blend of \ion{Si}{IV} $\lambda$4089 and \ion{O}{II} $\lambda$4089 lines. Other cases that we have also carefully checked are \ion{Si}{IV}$\lambda$4631, and \ion{Si}{II} $\lambda$6347, $\lambda$3856 since they could be blended with \ion{N}{II} $\lambda$4631, \ion{Mg}{II} $\lambda$6347, and \ion{O}{II} $\lambda$3856, respectively.

In the FASTWIND oxygen grid,  only lines from one ionization state were available (\ion{O}{II}), so we could not use the oxygen ionization equilibrium. Therefore, we adopted for the oxygen abundance analysis the same stellar parameters as in the silicon analysis, including microturbulence. Although \cite{ssimon10} used the microturbulence obtained from the \ion{O}{II} lines, we are more limited in the number of suitable spectral lines and decided to use the microturbulence derived from the silicon lines to avoid spurious results. 

 In Appendix~\ref{app} the results of the silicon and oxygen abundance analysis obtained for each line are detailed, where the uncertainties related to the errors in EWs ($\Delta\epsilon$), the standard weighted deviation in the line-by-line abundances ($\Delta\epsilon(\sigma)$) and the uncertainties associated with the microturbulence ($\Delta\epsilon(\xi)$) are indicated.

We observed no clear dependence between the derived abundances and $T_{\rm eff}$. However, due to the strong sensitivity of some line EWs to temperature we check the effect of varying $T_{\rm eff}$ in $\pm$ 500 K (taking the corresponding gravity variation into account).  In the case of silicon, the presence of two ionization stages already restricts the abundance range. 
 In contrast, in the case of oxygen, only lines from one ionization state were available, and therefore the derived abundance is quite sensitive to small changes in temperature. \cite{ssimon10} tested this effect in a sample of early dwarf  B  stars. He showed that the minimum variation in oxygen abundance occurs for $T_{\rm eff}$ $\sim$ 27000 K, increasing towards higher and lower effective temperatures.  We find the same behavior in our study.

\section{Results and discussion}\label{sect4}

\subsection{Chemical composition}\label{sect41}

Table~\ref{table3b} gives the microturbulence ($\xi$) and final abundance values (where $\epsilon_{x}$= 12+log(X/H)) derived for our stellar sample. It also shows the uncertainties in the derived abundances associated with the uncertainties in microturbulence ($\Delta\epsilon(\xi)$), the standard weighted deviation in the line-by-line abundances ($\Delta\epsilon(\sigma)$) and the impact that a modification of $T_{\rm eff}$ by $\pm$ 500 K and log $g$ by $\pm$ 0.1 dex have on the derived abundances ($\Delta\epsilon(T_{eff},log g)$).

Figure~\ref{fig10} represents the derived silicon and oxygen abundances as a function of Galactic longitude for our sample of stars.
We find chemical homogeneity for both elements, and derive mean silicon and oxygen values of 7.53 dex and 8.65 dex, respectively, that represent the mean abundance values of Cygnus OB2. The standard deviation for the silicon abundance is 0.08 dex and for the oxygen abundance 0.12 dex. The intrinsic uncertainties (related to the microturbulence, dispersion line-by-line and stellar parameters) are on average of the order of the dispersion in abundances (around 0.1 dex for both silicon and oxygen). 
The larger dispersion obtained for oxygen, in comparison with silicon, is partially due to 
the fact that only one ionization state is available for the oxygen analysis, while two ionization states are available for silicon (\ion{Si}{II-III} or \ion{Si}{III-IV}, depending on the temperature). 

Thus our results indicate a homogeneous composition for our stellar sample, without evidence of a dependence on the Galactic longitude (which, according to \cite{com12} or \cite{berlanas17} correlates with stellar age). Appendix~\ref{app2} shows the individual spectra with the best fitting FASTWIND model including \ion{H}{I}, \ion{He}{I-II}, \ion{Si}{II-III-IV} and \ion{O}{II}.

\subsection{Self-enrichment scenario}\label{sect43}

\begin{figure}[t!]
\centering
\includegraphics[width=8.0cm]{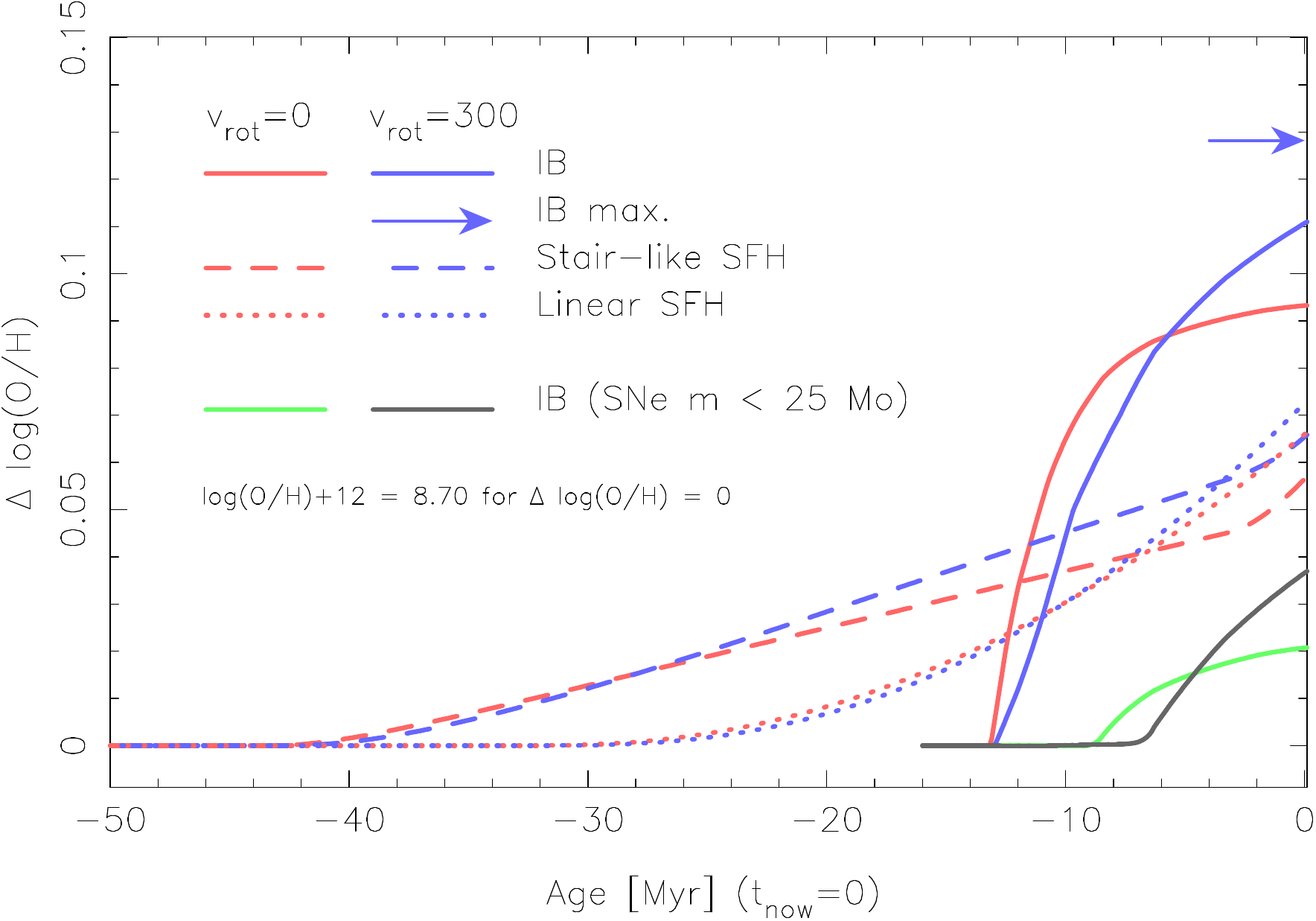}
\caption{Time evolution of the $\log \mathrm{O}/\mathrm{H}$ variation obtained from solar metallicity chemical evolution models, using three different SFHs (an IB and two extended SFH, both with a difference of a factor 6 between the actual SF value and the value 30 Myrs ago, but with different shapes), and two SNe cases: one where all massive stars produce a SNe explosion and another where stars more massive than 25 M$_{\odot}$ suffer a direct collapse to black holes (where only the IB case is shown). In all cases a star formation efficiency of 10\% is assumed. The arrow in the top right shows the maximum enrichment reached by the three SFHs.}
\label{figOH}
\end{figure}

In order to see whether we can provide some constraints on the metallicity enrichment within star-forming clouds in spite of the small variations obtained for the oxygen and silicon abundances, we estimate in this section the enrichment produced by stellar winds and SNe that may have polluted the region in the last 30 Myrs. This is roughly the time span during which a cluster would contain stars hotter than 20000 K according the  evolutionary tracks with and without rotation by \citealt{ekstrom12} and \citealt{LC18}. 

In the following we assume an Initial Mass Function between 0.1 and 120 M$_{\odot}$ with a slope $\alpha = -2.3$, as in \cite{com16}. We also assume three star formation histories (SFH): (1) An Instantaneous Burst (IB) having taken place 16 Myrs ago (corresponding roughly to the age when the number of red supergiants (RSGs) is maximum along the IB evolution for solar metallicity models by \citealt{ekstrom12}). This scenario shows the enrichment due to stars more massive than 15 $\mathrm{M}_\sun$ and it is used as an estimate of the maximum possible enrichment speed (see below). (2) A stair-like SFH with a constant value $\Phi_1$ in the last 6 Myrs plus a constant value $\Phi_2 = \frac{1}{6}\Phi_1$ for ages larger than 6 Myrs \citep[see again][]{com16}; in our case we have extended $\Phi_2$ up to 46 Myrs which is the maximum age where RSGs (defined as stars with $\log L/L_{\odot} \gtrsim 4$ and $T_\mathrm{eff} \lesssim 4000 \mathrm{K}$) would be present in a system, hence the lower limit where \cite{com16} SFH inference is applicable. And (3) a linearly increasing SFH from $\Phi(t_\mathrm{now} - 30 \mathrm{Myrs}) =  \frac{1}{6} \Phi(t_\mathrm{now})$ to  $\Phi(t_\mathrm{now})$, where $t_\mathrm{now}$ is the present moment and which is a linear approximation of the previous case; this SFH extends down to 35 Myrs (larger ages produce negative values). 
We have used yields and stellar lifetimes from \cite{LC18}, solar metallicity evolutionary models\footnote{Available at http://orfeo.iaps.inaf.it.} (see also \citealt{prantzos18}), which include three initial rotational velocities $v_\mathrm{rot}$ (0, 150 and 300 km s$^{-1}$), and two SNe scenarii: a first one where all stars produce SNe explosions, and a second one where stars more massive than 25 $\mathrm{M}_{\odot}$ suffer a direct collapse to black holes. In all cases the enrichment due to stellar winds in the pre-SNe phase (hydrostatic enrichment) has been taken into account, although we have assumed that the hydrostatic enrichment is released to the ISM at the end of the evolution of the star together with the SNe contribution. However, stellar winds produce a minimal effect in the abundance, when defined as a ratio (see below). Note that we have not consider variation of the yields with the metallicity; this choice is justified a posteriori given the small variation of the total abundance. For all SFHs we have followed the evolution of the total, H, O and Si masses released to the ISM as the stellar populations evolves. As usual in synthesis models computations, the result are normalized to total mass of ever formed stars (i.e. $\int^{t_\mathrm{now}}_{t_\mathrm{SFH\,begins}} \Phi(t)\, \mathrm{d}t = 1 \mathrm{M}_\sun$). Such choice allows to compare the different SFHs self consistently since the total (normalized\footnote{Actually, mean values of the underling possible distribution \citep{CL06}. Note that to make choice of an absolute value of the total mass, also implies to provide an study of the related (sampling) distribution, which is outside the scope of this paper. The current modeling is based in \cite{Cetal01,CM02} studies, and we also refer to these papers the study of sampling distributions in the context of chemical evolution.}) mass of the ejected material is equal in all cases. The only differences among the SFHs are the speed in which such material is released, being the IB case the faster case since all stars have equal ages and the only delay is due to the stellar evolution lifetimes.

In order to obtain the chemical enrichment, we assumed a close system with a global efficiency of 10\%, being the efficiency the ratio between total amount of gas ever transformed into stars (the integral over time of the SFH up to present time) and the total mass of the system. 
We note that efficiency is enough to follow the enrichment evolution of ISM if sampling effects are neglected, since the absolute value of the total mass cancels out when the O/H ratio is computed.
As a final note, the initial metallicity assumed in computations and given by the tracks, $\log(\mathrm{O}/\mathrm{H}) + 12 = 8.70$ dex, is larger than the possible oxygen abundance of Cyg OB2. As a result, rather than the nominal values of the abundance, we are interested in the abundance variation and how the slope of the chemical enrichment varies with time and, in particular,  in the last 20-25 Myrs when the present O-type stars were formed.

The result of the chemical evolution variation is shown in Fig. \ref{figOH} for the case of O/H (results for Si/H are similar, although with a slightly narrower range of variation). We note that since we are working with the O/H ratio, the enrichment is driven by the populations where (a) the O/H ratio of the ejected material differs significantly from the initial one, and (b) their overall contribution is large in comparison with the other stars which release material to the ISM. Such situation is only meet by the SNe contribution. Actually, there is almost no abundance variation in the IB case where stars more massive than 25 $\mathrm{M}_\sun$ collapse to black holes (red and blue solid lines) up to the age when star less massive than 25 $\mathrm{M}_\sun$ explode as SNe (green and black solid lines). As a secondary effect, the possible enrichment saturates with time, since the O/H ratio of the ejected material approaches to the initial value when the initial stellar mass decreases. Actually, all models reach a steady state value of the enrichment when extended to larger (future) ages, provided the SFHs stops at $t_\mathrm{now}$. Such steady state is reached around 25 Myrs after the onset of the burst in the IB case, and it is shown in the figure by an horizontal arrow for the extreme case of maximum enrichment. We stress that it does not means that there is no more oxygen released to the ISM, but that such oxygen enrichment is compensated by the hydrogen enrichment. The figure shows that differences between evolutionary models with and without rotation is not larger than 0.02 dex (5\%) beyond present-day accuracies in the spectral analysis. On the other hand, the scenario where all stars produce a SNe explosion (red and blue solid lines, with a maximum possible enrichment of 0.14 dex)  and that where only stars less massive than 25 $M_{\odot}$ produce a SNe (green and black solid lines, with a maximum enrichment of 0.05 dex) could be distinguished with techniques like those used in the present work, provided that we extend our sample and/or increase the SNR of the spectra. The only quantity we can constrain is the star forming efficiency. With values significantly larger than the one adopted here (10$\%$) we would see metallicity variations in our sample. Finally, if we would  firmly constrain the age of the different stars, it would be much easier to discriminate between different SFHs. Gaia DR2 and future DR3 data may be unvaluable to this aim. However, given that the different scenarios produce metallicity variations which are similar to the uncertainties in abundance determination of the current analysis, we conclude that the massive star forming activity during the last 20-25 Myr in Cygnus OB2 has not caused any detectable increase in its O/H and Si/H abundance,
within the limitations of the present study, in agreement with the findings in Orion by \cite{ssimon10} or  \cite{biazzo11a}.

\subsection{The Galactic context}\label{sect411}

Our abundance results are in agreement with the values obtained in Orion OB1 for a sample of early-B type stars \citep{ssimon10}, and also with those obtained for a sample of unevolved early B-type stars in the solar neighborhood \citep{pryzbilla08}. Both studies result in a small dispersion of abundances, indicating a homogeneity of oxygen and silicon among B-type stars. 

\begin{figure}[t!]
\centering 
\includegraphics[width=9.0cm,height=4.5cm ]{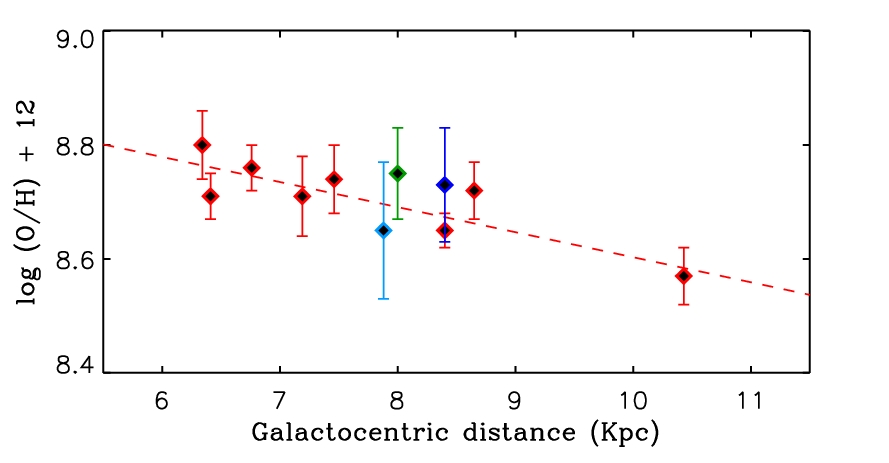}
\caption{ Spatial distribution of the oxygen abundance as a function of Galactocentric distances for Galactic \ion{H}{II} regions (red dots), the mean value obtained for  Orion OB1 (dark blue dot), the mean value obtained for the solar vicinity (green dot)  and  the mean value obtained for Cygnus OB2 (light blue dot). The red dashed line represents the least-squares linear fits to the \ion{H}{II} regions. The Sun is assumed at 8 kpc.}
\label{rad_ab_grad} 
\end{figure}

Regarding oxygen, we find a mean abundance value slightly below the one obtained in both Orion OB1 and the solar vicinity. In order to check whether they are consistent with the oxygen radial abundance gradient  of the Milky Way, we have compared these values with the work of \cite{esteban05}. They presented homogeneous data of eight \ion{H}{II} regions with Galactocentric distances (R$_{\rm G}$) between 6.3 and 10.4 kpc using recombination lines and assuming the Sun at 8 kpc from the Galactic center. Figure~\ref{rad_ab_grad} shows the spatial distribution of the oxygen abundance as a function of R$_{\rm G}$ for the considered Galactic \ion{H}{II} regions and the mean values obtained for Orion OB1, the solar vicinity, and Cygnus OB2.
All of them are consistent with the observed abundance gradient derived from \ion{H}{II} regions in the Milky Way and its dispersion. Although one would expect a large abundance value for Cygnus OB2 due to its shorter Galactocentric distance, the average value is consistent within uncertainties with the Orion OB1, the solar neighborhood and the \ion{H}{II} regions. Our conclusion is that the observed scatter associated with individual measurements of Milky Way \ion{H}{II} regions at a given Galactocentric distance is at least of the same order than within a cluster or association.

\section{Summary and conclusions}\label{sect5}

We have presented the spectroscopic analysis of eight late O, early B stars in Cygnus OB2 using the curve-of-growth method to derive silicon and oxygen abundances. We have used high S/N, high resolution spectra obtained with ISIS at the WHT, and FASTWIND stellar atmosphere models.
Our targets are early-type slow rotators ($v$ sin $i <$ 80 km~s$^{-1}$) with temperatures below 34 000 K selected from a list of 57 spectroscopically classified OB stars in Cygnus OB2. They have late O or early B spectral type so that we can use their \ion{Si}{II-III-IV} lines for our abundance analysis, and are distributed all across the association area. According to Gaia data there is no evidence of a peculiar behavior indicating a runaway origin of our stellar sample. Thus, the silicon and oxygen abundances of the targeted stars are considered to reflect those of the cloud out of which the stars formed.

The abundance analysis was based on the EW of silicon and oxygen lines and the Balmer line profiles. We first obtain the fundamental parameters ($T_{\rm eff}$ and log $g$) by comparing the EW ratio of \ion{Si}{II-III} or \ion{Si}{II-IV}, depending on the temperature of each star, and the observed wings of H Balmer lines with FASTWIND models. Then, using these parameters,  the  abundance was computed for different values of microturbulence and for each line considered. Finally, we determined the final abundance value considering the microturbulence that, for all lines, gives the same abundance value.

Both silicon and oxygen analyses suggest a homogeneous stellar chemical  composition in Cygnus OB2. We find a mean abundance value of 7.53 dex for silicon and 8.65 dex for oxygen, with a standard deviation of 0.08 dex and 0.12 dex, respectively.

We roughly estimate the contribution to oxygen enrichment of SNe that may have exploded in the region, concluding that the original cloud from which stars formed in the last 20-25 Myr in Cygnus OB2  
had a homogeneous metallicity and has not been significantly enriched during its lifetime as a consequence of its star forming activity in agreement with the findings in Orion OB1 and the solar vicinity. All these studies are consistent with the spatial abundance distribution for \ion{H}{II} regions in the Milky Way. The oxygen abundance scatter based on \ion{H}{II} regions at a given Galactocentric distance is of the same order than within a cluster or association. Therefore, abundance gradients at the local scale, like Cygnus OB2 or Orion OB1, are not expected. We estimate that the effect of self-enrichment by stellar winds and SNe is beyond the accuracy of our analyses, except for the star forming efficiency that cannot be significantly larger than the value of 10$\%$ adopted here. 

Our computations illustrate the difficulties of making quantitative predictions on the level of enrichment expected in a massive association like Cygnus OB2, and show in any case that such level should be small as compared to the accuracies currently attainable.

\begin{acknowledgements}

We thank the referee for useful and valuable comments that helped improve this paper. We also thank F. Najarro, M.A. Urbaneja, and D.J. Lennon for helpful discussions and comments. We thank J. Ma\'iz Apell\'aniz for advice on the use of MGB and L. López-Martín for advice on spectral data reduction. We acknowledge financial support from the Spanish Ministry of Economy and Competitiveness (MINECO) under the grants AYA 2015-68012-C2-01 and Severo Ochoa SEV-2015-0548 and the Gobierno de Canarias under grant ProID-2017010115. AP acknowledges support from  Sonderforschungsbereich  SFB  881 ‘The Milky Way System’ (subproject B5) of the German Research Foundation (DFG).

\end{acknowledgements}

%
%

\def\bibname{References}

\begin{appendix}

\section{Rotational velocities}\label{app_vsini}
Table~\ref{vsini} shows the whole list of stars observed in July 2014 and August 2012 at the WHT. Names, spectral types and the derived rotational velocities are indicated. 

  \begin{table*}[p!]
	\centering
	\caption{List of stars observed in July 2014 and August 2012 at the WHT. Names, spectral types, coordinates (J2000.0) and the derived projected rotational velocities are indicated. Asterisks indicate possible SB2 stars, whose spectral types are refereed to the primary component.}
	\label{vsini}
		\begin{tabular}{lcccc}
		\hline   
		\hline\\[-1.8ex]
\small{Target} & \small{SpT} & \small{Ref.}&\small{Coordinates ($^{\circ}$)} & \small{$v$ sin $i$ (km~s$^{-1}$)} \\    	
\cline{1-5}\\[-1.5ex]

\small{J20332346+4109130} & \small{O6IV((f))} & \small{G16}& \small{308.35 +41.15} & \small{100} \\
\small{J20331326+4113287} & \small{O6V} & \small{K07}& \small{308.30 +41.22} & \small{ 240} \\ 
\small{HD195213 } & \small{O7} & \small{M55}& \small{307.13 +40.82} & \small{90}\\ 
\small{J20334086+4130189} & \small{O7V} & \small{K07}& \small{308.42 +41.50} & \small{55} \\
\small{J20310019+4049497} & \small{O7V((f))} & \small{H03}& \small{307.75 +40.83} & \small{95}\\
\small{J20341350+4135027} & \small{O7.5(n)((f))z} & \small{G16} & \small{308.56 +41.58}& \small{190} \\	
\small{J20323857+4125137} & \small{O7.5IV(n)} & \small{G16}& \small{308.16 +41.42} & \small{195} \\
\small{J20315961+4114505} & \small{O7.5Vz} & \small{G16} & \small{307.99 +41.25} & \small{185}\\
\small{J20342959+4131455} & \small{O7.5V(n)((f))*} & \small{G16}& \small{308.62 +41.53} & \small{325} \\
\small{J20293563+4024315} & \small{O8IIIz} & \small{B18} & \small{307.39 +40.41} & \small{60}\\
\small{J20332674+4110595} & \small{O8.5Vz} & \small{G16} & \small{308.36 +41.18}& \small{95} \\
\small{J20340486+4105129} & \small{O8.5V} & \small{C12}& \small{308.52 +41.08} & \small{45} \\
\small{J20333030+4135578} & \small{O8V} & \small{M91}& \small{308.37 +41.59} & \small{120} \\	
\small{J20323486+4056174} & \small{O8V} & \small{N08} & \small{308.14 +40.94} & \small{130}\\
\small{J20325002+4123446} & \small{O8V} & \small{K07} & \small{308.21 +41.39} & \small{85}\\
\small{J20325919+4124254} & \small{O8V} & \small{K07} & \small{308.25 +41.41} & \small{155}\\		
\small{J20330292+4117431} & \small{O8V} & \small{K07} & \small{308.26 +41.29} & \small{175}\\	
\small{J20331803+4121366} & \small{O8V} & \small{K07}& \small{308.32 +41.36} & \small{110} \\
\small{J20313749+4113210} & \small{O9:} & \small{M55}& \small{307.91 +41.22} & \small{115} \\
\small{J20311055+4131535} & \small{O9V} & \small{M55} & \small{307.79 +41.53}& \small{235} \\	
\small{J20301839+4053466} & \small{O9V} & \small{C12}& \small{307.57 +40.89} & \small{55}\\	
\small{J20321656+4125357} & \small{O9V} & \small{K07} & \small{308.07 +41.43}& \small{215} \\
\small{J20332101+4117401} & \small{O9V} & \small{K07} & \small{308.34 +41.29}& \small{130} \\
\small{J20323033+4034332} & \small{O9.5IV} & \small{H03} & \small{308.13 +40.57} & \small{120}\\
\small{J20293480+4120089} & \small{O9.5V} & \small{C12} & \small{307.39 +41.33}  & \small{80}\\
\small{J20341605+4102196} & \small{O9.5V} & \small{C12}& \small{308.57 +41.04} & \small{175} \\		
\small{J20291617+4057372} & \small{O9.7III} & \small{B18}& \small{307.32 +40.96} & \small{100}\\
\small{J20295701+4109538} & \small{O9.7III} & \small{New}& \small{307.49 +41.16} & \small{70}\\	
\small{J20335952+4117354} & \small{O9.7V(n)} & \small{G16}& \small{308.49 +41.29} & \small{335}\\
\small{J20323968+4050418} & \small{B0II} & \small{B18} & \small{308.16 +40.84}& \small{75}\\
\small{J20272428+4115458} & \small{B0IV} & \small{New}& \small{306.85 +41.26} & \small{20} \\
\small{J20340601+4108090} & \small{B0V} & \small{C12}& \small{308.52 +41.13} & \small{145} \\		
\small{J20331050+4122224} & \small{B0V} & \small{K07} & \small{308.29 +41.37}& \small{110} \\	
\small{J20305552+4054541} & \small{B0V} & \small{N08}& \small{307.73 +40.91} & \small{180}\\
\small{J20312210+4112029} & \small{B0.2III} & \small{C12}& \small{307.84 +41.20}  & \small{245}\\
\small{J20292449+4052599} & \small{B0.2IV} & \small{C12}& \small{307.35 +40.88}  & \small{75}\\
\small{J20321568+4046170} & \small{B0.2IV} & \small{C12} & \small{308.06 +40.77} & \small{45}\\
\small{J20330526+4143367} & \small{B0.5III} & \small{B18}  &\small{308.27 +41.73}& \small{115}\\
\small{J20314605+4043246} & \small{B0.5IV} & \small{H03} & \small{307.94 +40.72} & \small{60}\\
\small{J20282772+4104018} & \small{B0.5V} & \small{C12}& \small{307.11 +41.07}   & \small{305}\\
\small{J20294666+4105083} & \small{B0.5V(n)} & \small{H03}& \small{307.44 +41.08} & \small{140} \\
\small{J20340435+4108078} & \small{B1III} & \small{C12} & \small{308.52 +41.13}& \small{310} \\	
\small{BD+40 4208 } & \small{B1V}  & \small{B18}& \small{307.71 +40.74}& \small{165} \\
\small{J20293473+4020381} & \small{B1V*} & \small{C12}& \small{307.39 +40.34}& \small{330}\\
\small{J20303297+4044024} & \small{B1V} & \small{C12}& \small{307.64 +40.73} & \small{195}\\
\small{J20303833+4010538} & \small{B1V} & \small{C12}& \small{307.66 +40.18} & \small{255}\\
\small{J20313338+4122490} & \small{B1V} & \small{K07}& \small{307.89 +41.38} & \small{175}\\
\small{J20314341+4100021} & \small{B1V*} & \small{B18} & \small{307.93 +41.00}  & \small{370}\\
\small{J20273982+4040384} & \small{B1V} & \small{C12}& \small{306.91 +40.68} & \small{175} \\
\small{J20315898+4107314} & \small{B1V*} & \small{B18} & \small{307.99 +41.12}  & \small{325}\\		
\small{J20303297+4044024} & \small{B1V} & \small{C12}& \small{307.64 +40.73} & \small{255} \\		
\small{BD+40 4210} & \small{B2Ia} & \small{New} & \small{307.77 +40.52}& \small{30} \\
\small{BD+40 4193 } & \small{B2V} & \small{New}& \small{307.31 +40.68}& \small{70}\\
\small{J20322734+4055184} & \small{B2V} & \small{H03} & \small{308.11 +40.92}& \small{200}\\ 		
\small{J20354703+4053012} & \small{B2V} & \small{C12} & \small{308.94 +40.88}& \small{215} \\	
\small{CCDM J20323+4152AB}& \small{B9V} & \small{B18}& \small{308.09 +41.87} & \small{110} \\	
\small{J20315984+4120354} & \small{A} & \small{B18} & \small{307.99 +41.34}  & \small{55}\\

\hline
\hline\\[-1.5ex]		 
		 		 		 
		\end{tabular}
		\tablefoot{Spectral type source: B18) \cite{berlanas17}, C12) \cite{com12}, G16) \cite{maiz16}, H03) \cite{hanson03}, K07) \cite{kiminki07}, M91) \cite{mt91}, N08) \cite{neg08}, M55) \cite{morgan55}, New) This work. }
\end{table*}

\section{Individual Targets}\label{app}

Tables~\ref{target1}-\ref{target8} show the detailed results of the silicon and oxygen abundance analysis line-by-line for each star of the sample. Abundances ($\xi(x)$) are given in units of 12 + log(X/H). We also indicate the final weighed mean value ($\bar{\epsilon}$) and the uncertainties related to the errors in EW measurements ($\Delta\epsilon$), the standard weighed deviation ($\Delta\epsilon(\sigma)$) and the uncertainties associated with the microturbulence ($\Delta\epsilon(\xi)$).

  \begin{table}[p!]
	\centering
	\caption{Detailed results of the silicon and oxygen abundance analysis line-by-line for the star J20314965+4128265  ($T_{\rm eff}$=33000 K, log $g$=4.0 dex and $\xi(Si)$=$\xi(O)$=3.0 km~s$^{-1}$).}
	\label{target2}
		\begin{tabular}{lcc}
		\hline   
		\hline\\[-1.8ex]
        \small{Line} & \small{EW (m$\AA$)} & \small{$\epsilon_{Si}\pm \Delta\epsilon_{Si}$ (dex)} \\    	
   		 \cline{1-3}\\[-1.5ex]
		 \small{\ion{Si}{III} 4552} & \small{129$\pm$20} & \small{7.83$\pm$0.32} \\ 
		  \small{\ion{Si}{III} 4567} & \small{103$\pm$13} & \small{7.78$\pm$0.22} \\
		  \small{\ion{Si}{III} 4574} & \small{55$\pm$6} & \small{7.72$\pm$0.13} \\
		  \small{\ion{Si}{III} 5739} & \small{70$\pm$20} & \small{7.60$\pm$0.17} \\
		  \small{\ion{Si}{IV} 4116} & \small{171$\pm$41} & \small{7.61$\pm$0.36}  \\ 
		  \small{\ion{Si}{IV} 4631} & \small{99$\pm$9} & \small{7.63$\pm$0.12}  \\        
		  \small{\ion{Si}{IV} 6701} & \small{68$\pm$6} & \small{7.63$\pm$0.12}  \\  
		 \cline{3-3}\\[-1.5ex]
		 \small{} & \small{} & \small{$\bar{\epsilon_{Si}}$=7.67} \\
		 \small{} & \small{} &\small{$\Delta\epsilon_{Si}(\sigma)$=0.06, $\Delta\epsilon_{Si}(\xi)$=0.08}\\    	    
			
        \cline{1-3}\\[-2.0ex] 
         \cline{1-3}\\[-1.8ex] 		
         \small{Line} & \small{EW (m$\AA$)} & \small{$\epsilon(O)\pm \Delta\epsilon_{O}$ (dex)}  \\    	
   		 \cline{1-3}\\[-1.5ex] 
		  \small{\ion{O}{II} 4317} & \small{66$\pm$15}& \small{8.62$\pm$0.24} \\
   		 \small{\ion{O}{II} 4319} & \small{74$\pm$15}& \small{8.77$\pm$0.26} \\
		 \small{\ion{O}{II} 4366} & \small{88$\pm$15}& \small{8.95$\pm$0.26} \\
		 \small{\ion{O}{II} 4414} & \small{81$\pm$15}& \small{8.47$\pm$0.24} \\
		 \small{\ion{O}{II} 4416} & \small{91$\pm$15}& \small{8.86$\pm$0.24} \\
		 \small{\ion{O}{II} 4638} & \small{72$\pm$15}& \small{8.94$\pm$0.31}\\
		 \small{\ion{O}{II} 4661} & \small{88$\pm$15}& \small{9.01$\pm$0.28}\\
		\small{\ion{O}{II} 4676} & \small{65$\pm$15}& \small{8.76$\pm$0.32}\\
		 \small{\ion{O}{II} 4072}& \small{102$\pm$15}& \small{8.95$\pm$0.31}\\
		 \small{\ion{O}{II} 4076}& \small{101$\pm$15}& \small{8.72$\pm$0.13}\\
		  \small{\ion{O}{II} 4189}& \small{56$\pm$15}& \small{8.45$\pm$0.30}\\
		 \cline{3-3}\\[-1.5ex] 
		 \small{} &\small{}& \small{$\bar{\epsilon_{O}}$=8.75}\\
		  \small{} &  \small{} &\small{$\Delta\epsilon_{O}(\sigma)$=0.07, $\Delta\epsilon_{Si}(\xi)$=0.02}\\ 		
		\hline
		\hline\\[-1.5ex]
				
		\end{tabular}
\end{table}			

  \begin{table}[p!]
	\centering
	\caption{Detailed results of the silicon and oxygen abundance analysis line-by-line for the star J20295701+4109538  ($T_{\rm eff}$=32000 K, log $g$=3.6 dex and $\xi(Si)$=$\xi(O)$=7.5 km~s$^{-1}$).}
	\label{target1}
		\begin{tabular}{lcc}
		\hline   
		\hline\\[-1.8ex]
         \small{Line} & \small{EW (m$\AA$)} & \small{$\epsilon_{Si}\pm \Delta\epsilon_{Si}$ (dex)} \\    	
   		 \cline{1-3}\\[-1.5ex]
		\small{\ion{Si}{III} 4552} & \small{138$\pm$15} & \small{7.58$\pm$0.16} \\ 
		\small{\ion{Si}{III} 4567} & \small{108$\pm$15} & \small{7.60$\pm$0.17} \\         
		\small{\ion{Si}{IV} 4116} & \small{241$\pm$15} & \small{7.63$\pm$0.17} \\ 
		\small{\ion{Si}{IV} 4212} & \small{81$\pm$15} & \small{7.66$\pm$0.22}  \\ 
		 \cline{3-3}\\[-1.5ex]
		\small{} & \small{} & \small{$\bar{\epsilon_{Si}}$=7.61} \\   
		\small{} & \small{} & \small{$\Delta\epsilon_{Si}(\sigma)$=0.09, $\Delta\epsilon_{Si}(\xi)$=0.12  } \\ 	    	
        \cline{1-3}\\[-2.0ex] 
        \cline{1-3}\\[-1.8ex] 		
         \small{Line} & \small{EW (m$\AA$)} & \small{$\epsilon(O)\pm \Delta\epsilon_{O}(\sigma)$ (dex)} \\    	
   		 \cline{1-3}\\[-1.5ex] 
		  \small{\ion{O}{II} 4641} & \small{125$\pm$15}& \small{8.74$\pm$0.14}\\
		 \small{\ion{O}{II} 4076}& \small{130$\pm$15}& \small{8.75$\pm$0.15}\\
		 \cline{3-3}\\[-1.5ex] 
		 \small{} & \small{}& \small{$\bar{\epsilon_{O}}$=8.74}\\ 
		 \small{} &	\small{}& \small{$\Delta\epsilon_{O}(\sigma)$=0.10, $\Delta\epsilon_{Si}(\xi)$=0.07}\\
	 	
		\hline
		\hline\\[-1.5ex]		
		\end{tabular}
\end{table}

  \begin{table}[p!]
	\centering
	\caption{Detailed results of the silicon and oxygen abundance analysis line-by-line for the star J20272428+4115458 ($T_{\rm eff}$=30000 K, log $g$=3.9 dex and  $\xi(Si)$=$\xi(O)$=5.2 km~s$^{-1}$). }
	\label{target3}
		\begin{tabular}{lcc}
		\hline   
		\hline\\[-1.8ex]	
         \small{Line} & \small{EW (m$\AA$)} & \small{$\epsilon_{Si}\pm \Delta\epsilon_{Si}$ (dex)} \\    	
   		 \cline{1-3}\\[-1.5ex]   	   	
\small{\ion{Si}{III} 4552} & \small{167$\pm$11} & \small{7.41$\pm$0.14} \\
\small{\ion{Si}{III} 4567} & \small{152$\pm$10} & \small{7.53$\pm$0.13} \\
\small{\ion{Si}{III} 4574} & \small{83$\pm$8} & \small{7.43$\pm$0.12} \\
\small{\ion{Si}{III} 4539} & \small{105$\pm$10} & \small{7.41$\pm$0.16} \\ 
\small{\ion{Si}{IV} 4116} & \small{145$\pm$20} & \small{7.41$\pm$0.29} \\        
\small{\ion{Si}{IV} 4212} & \small{44$\pm$9} & \small{7.57$\pm$0.23} \\ 
\small{\ion{Si}{IV} 4654} & \small{73$\pm$14} & \small{7.29$\pm$0.25}  \\		  
		 \cline{3-3}\\[-1.5ex]
         \small{} & \small{} & \small{$\bar{\epsilon_{Si}}$=7.45} \\	
         \small{} & \small{} & \small{$\Delta\epsilon_{Si}(\sigma)$=0.06, $\Delta\epsilon_{Si}(\xi)$=0.09}\\           	  	    	
          \cline{1-3}\\[-2.0ex] 
          \cline{1-3}\\[-1.8ex] 	
          \small{Line} & \small{EW (m$\AA$)} & \small{$\epsilon(O)\pm \Delta\epsilon_{O}$ (dex)}\\    	
   		 \cline{1-3}\\[-1.5ex] 
\small{\ion{O}{II} 4317} & \small{98$\pm$15}& \small{8.57$\pm$0.19} \\
\small{\ion{O}{II} 4319} & \small{093$\pm$16}& \small{8.58$\pm$0.24} \\
\small{\ion{O}{II} 4366} & \small{102$\pm$15}& \small{8.68$\pm$0.23} \\
\small{\ion{O}{II} 4414} & \small{125$\pm$15}& \small{8.44$\pm$0.19} \\
\small{\ion{O}{II} 4416} & \small{126$\pm$15}& \small{8.74$\pm$0.20} \\
\small{\ion{O}{II} 4638} & \small{93$\pm$15}& \small{8.77$\pm$0.26}\\
\small{\ion{O}{II} 4661} & \small{113$\pm$15}& \small{8.84$\pm$0.24}\\
\small{\ion{O}{II} 4676} & \small{77$\pm$15}& \small{8.49$\pm$0.27}\\
\small{\ion{O}{II} 4076}& \small{132$\pm$15}& \small{8.66$\pm$0.25}\\
\small{\ion{O}{II} 4132}& \small{56$\pm$12}& \small{8.93$\pm$0.23}\\
\small{\ion{O}{II} 4185}& \small{61$\pm$13}& \small{8.41$\pm$0.23}\\
\small{\ion{O}{II} 4189}& \small{90$\pm$15}& \small{8.58$\pm$0.21}\\		 
		 \cline{3-3}\\[-1.5ex] 
        \small{} & \small{}& \small{$\bar{\epsilon_{O}}$=8.63} \\	
        \small{} & \small{}& \small{$\Delta\epsilon_{O}(\sigma)$=0.06, $\Delta\epsilon_{O}(\xi)$=0.12} \\        	  	
		\hline
		\hline\\[-1.5ex]
			
		\end{tabular}
\end{table}

  \begin{table}[p!]
	\centering
	\caption{Detailed results of the silicon and oxygen abundance analysis line-by-line for the star J20334610+4133010 ($T_{\rm eff}$=30000 K, log $g$=3.2 dex and  $\xi(Si)$=$\xi(O)$=17.0 km~s$^{-1}$). }
	\label{target4}
		\begin{tabular}{lcc}
		\hline   
		\hline\\[-1.8ex]	
         \small{Line} & \small{EW (m$\AA$)} & \small{$\epsilon_{Si}\pm \Delta\epsilon_{Si}$ (dex)} \\    	
   		 \cline{1-3}\\[-1.5ex]   	   	
\small{\ion{Si}{III} 4552} & \small{188$\pm$21} & \small{7.42$\pm$0.12} \\
\small{\ion{Si}{III} 4567} & \small{148$\pm$16} & \small{7.49$\pm$0.09} \\
\small{\ion{Si}{III} 4574} & \small{72$\pm$15} & \small{7.57$\pm$0.15} \\
\small{\ion{Si}{IV} 4116} & \small{445$\pm$30} & \small{7.70$\pm$0.06} \\        
\small{\ion{Si}{IV} 4631} & \small{100$\pm$20} & \small{7.37$\pm$0.23} \\ 	  
		 \cline{3-3}\\[-1.5ex]
           \small{} & \small{} & \small{$\bar{\epsilon_{Si}}$=7.54} \\	
           \small{} & \small{} & \small{$\Delta\epsilon_{Si}(\sigma)$=0.02, $\Delta\epsilon_{Si}(\xi)$=0.04 } \\           	  	    	
          \cline{1-3}\\[-2.0ex] 
          \cline{1-3}\\[-1.8ex] 	
          \small{Line} & \small{EW (m$\AA$)} & \small{$\epsilon(O)\pm \Delta\epsilon_{O}$ (dex)}\\    	
   		 \cline{1-3}\\[-1.5ex] 
\small{\ion{O}{II} 4317} & \small{98$\pm$15}& \small{8.76$\pm$0.15} \\
\small{\ion{O}{II} 4366} & \small{97$\pm$20}& \small{8.69$\pm$0.20} \\
\small{\ion{O}{II} 4069}& \small{110$\pm$20}& \small{8.45$\pm$0.18}\\
\small{\ion{O}{II} 4076}& \small{126$\pm$20}& \small{8.43$\pm$0.17}\\
\small{\ion{O}{II} 4132}& \small{34$\pm$10}& \small{8.49$\pm$0.15}\\
	 
		 \cline{3-3}\\[-1.5ex] 
        \small{} & \small{}& \small{$\bar{\epsilon_{O}}$=8.57} \\	
        \small{} & \small{}& \small{$\Delta\epsilon_{O}(\sigma)$=0.07, $\Delta\epsilon_{O}(\xi)$= 0.10} \\        	  	
		\hline
		\hline\\[-1.5ex]
			
		\end{tabular}
\end{table}

  \begin{table}[p!]
	\centering
	\caption{Detailed results of the silicon and oxygen abundance analysis line-by-line for the star J20314605+4043246 ($T_{\rm eff}$=30000 K, log $g$=4.0 dex and $\xi(O)$=$\xi(Si)$=5.6 km~s$^{-1}$). }
	\label{target5}
		\begin{tabular}{lcc}
		\hline   
		\hline\\[-1.8ex]
         \small{Line} & \small{EW (m$\AA$)} & \small{$\epsilon_{Si}\pm \Delta\epsilon_{Si}$ (dex)} \\    	
   		 \cline{1-3}\\[-1.5ex]
		\small{\ion{Si}{III} 4552} & \small{185$\pm$15} & \small{7.49$\pm$0.17} \\         
		\small{\ion{Si}{III} 4567} & \small{156$\pm$20} & \small{7.50$\pm$0.24} \\ 
		\small{\ion{Si}{III} 4574} & \small{97$\pm$10} & \small{7.52$\pm$0.14} \\
		\small{\ion{Si}{III} 5739} & \small{150$\pm$20} & \small{7.77$\pm$0.23} \\
		\small{\ion{Si}{IV} 4116} & \small{140$\pm$20} & \small{7.42$\pm$0.29}  \\  
		 \cline{3-3}\\[-1.5ex]
	    \small{} & \small{} & \small{$\bar{\epsilon_{Si}}$=7.54} \\ 
	    \small{} & \small{} & \small{$\Delta\epsilon_{Si}(\sigma)$=0.09, $\Delta\epsilon_{Si}(\xi)$=0.12} \\ 	      	
        \cline{1-3}\\[-2.0ex] 
        \cline{1-3}\\[-1.8ex] 	
         \small{Line} & \small{EW (m$\AA$)} & \small{$\epsilon(O)\pm \Delta\epsilon_{O}$ (dex)}\\    	
   		 \cline{3-3}\\[-1.5ex] 
		 \small{\ion{O}{II} 4366} & \small{111$\pm$20}& \small{8.74$\pm$0.29} \\
		\small{\ion{O}{II} 4452} & \small{83$\pm$15}& \small{9.04$\pm$0.24} \\
		\small{\ion{O}{II} 4661} & \small{91$\pm$20}& \small{8.55$\pm$0.34}\\
		\small{\ion{O}{II} 4676} & \small{104$\pm$20}& \small{8.79$\pm$0.33}\\
		\small{\ion{O}{II} 6641} & \small{63$\pm$15}& \small{9.14$\pm$0.28}\\
		\small{\ion{O}{II} 6721} & \small{48$\pm$15}& \small{8.54$\pm$0.29}\\
		\small{\ion{O}{II} 4076}& \small{137$\pm$20}& \small{8.67$\pm$0.33}\\
		\small{\ion{O}{II} 4132}& \small{56$\pm$15}& \small{8.90$\pm$0.30}\\
		\small{\ion{O}{II} 4189}& \small{104$\pm$20}& \small{8.71$\pm$0.29}\\
		 \cline{1-3}\\[-1.5ex] 
		\small{} & \small{Final value}& \small{$\bar{\epsilon_{O}}$=8.82} \\   
		\small{} & \small{}& \small{$\Delta\epsilon_{O}(\sigma)$=0.10, $\Delta\epsilon_{O}(\xi)$=0.12} \\ 			
		\hline
		\hline\\[-1.5ex]	
		
		\end{tabular}
\end{table}		

  \begin{table}[p!]
	\centering
	\caption{Detailed results of the silicon and oxygen abundance analysis line-by-line for the star J20292449+4052599 ($T_{\rm eff}$=29000 K, log $g$=3.7 dex and $\xi(O)$=$\xi(Si)$=6.8 km~s$^{-1}$). }
	\label{target6}
		\begin{tabular}{lcc}
		\hline   
		\hline\\[-1.8ex]
         \small{Line} & \small{EW (m$\AA$)} & \small{$\epsilon_{Si}\pm \Delta\epsilon_{Si}$ (dex)} \\    	
   		 \cline{1-3}\\[-1.5ex]
		\small{\ion{Si}{III} 4552} & \small{222$\pm$15} & \small{7.49$\pm$0.25} \\         
		\small{\ion{Si}{III} 4567} & \small{185$\pm$15} & \small{7.47$\pm$0.08} \\ 
		\small{\ion{Si}{III} 4574} & \small{112$\pm$15} & \small{7.48$\pm$0.19}  \\ 
		 \cline{3-3}\\[-1.5ex]
		\small{} & \small{} & \small{$\bar{\epsilon_{Si}}$=7.47} \\	
		\small{} & \small{} & \small{$\Delta\epsilon_{Si}(\sigma)$=0.07, $\Delta\epsilon_{Si}(\xi)$=0.10} \\			    	
        \cline{1-3}\\[-2.0ex] 
        \cline{1-3}\\[-1.8ex] 	
         \small{Line} & \small{EW (m$\AA$)} & \small{$\epsilon(O)\pm \Delta\epsilon_{O}$ (dex)} \\    	
   		 \cline{1-3}\\[-1.5ex] 
		 \small{\ion{O}{II} 4366} & \small{89$\pm$20}& \small{8.37$\pm$0.29} \\
		 \small{\ion{O}{II} 4076}& \small{145$\pm$20}& \small{8.59$\pm$0.30}\\
		 \small{\ion{O}{II} 4661} & \small{115$\pm$20}& \small{8.65$\pm$0.29}\\
		 \small{\ion{O}{II} 4676}& \small{88$\pm$20}& \small{8.44$\pm$0.31}\\
		 \cline{3-3}\\[-1.5ex] 
		 \small{} & \small{}& \small{$\bar{\epsilon_{O}}$=8.51} \\ 
		 \small{} & \small{}& \small{$\Delta\epsilon_{O}(\sigma)$=0.15, $\Delta\epsilon_{O}(\xi)$=0.02} \\		    	
		\hline
		\hline\\[-1.5ex]		
		
		\end{tabular}
\end{table}

  \begin{table}[p!]
	\centering
	\caption{Detailed results of the silicon and oxygen abundance analysis line-by-line for the star BD+40 4193 ($T_{\rm eff}$=19000 K, log $g$=3.8 dex and $\xi(O)$=$\xi(Si)$=5.9 km~s$^{-1}$). }
	\label{target7}
		\begin{tabular}{lcc}
		\hline   
		\hline\\[-1.8ex]
        \small{Line} & \small{EW (m$\AA$)} & \small{$\epsilon_{Si}\pm \Delta\epsilon_{Si}$ (dex)}  \\    	
   		 \cline{1-3}\\[-1.5ex]
		\small{\ion{Si}{II} 6347} & \small{107$\pm$20} & \small{7.45$\pm$0.27} \\    
		\small{\ion{Si}{III} 4552} & \small{113$\pm$15} & \small{7.47$\pm$0.22} \\         
		\small{\ion{Si}{III} 4567} & \small{86$\pm$11} & \small{7.40$\pm$0.19} \\ 
		\small{\ion{Si}{III} 4574} & \small{57$\pm$15} & \small{7.54$\pm$0.34} \\ 
		\small{\ion{Si}{III} 5739} & \small{43$\pm$15} & \small{7.40$\pm$0.39} \\ 
		 \cline{3-3}\\[-1.5ex]
   		\small{} & \small{} & \small{$\bar{\epsilon_{Si}}$=7.44} \\  
    	\small{} & \small{} & \small{$\Delta\epsilon_{Si}(\sigma)$=0.11, $\Delta\epsilon_{Si}(\xi)$=0.12}  \\    			
        \cline{1-3}\\[-2.0ex] 
        \cline{1-3}\\[-1.8ex] 	
         \small{Line} & \small{EW (m$\AA$)} & \small{$\epsilon(O)\pm \Delta\epsilon_{O}$ (dex)} \\    	
   		 \cline{1-3}\\[-1.5ex] 
		 \small{\ion{O}{II} 4366} & \small{37$\pm$15}& \small{8.84$\pm$0.40} \\
		 \small{\ion{O}{II} 4650} & \small{52$\pm$15}& \small{8.63$\pm$0.31}\\
		 \small{\ion{O}{II} 4661}& \small{34$\pm$15}& \small{8.86$\pm$0.41}\\
		 \small{\ion{O}{II} 4072}& \small{25$\pm$15}& \small{8.38$\pm$0.42}\\
		 \cline{3-3}\\[-1.5ex] 
		 \small{} & \small{}& \small{$\bar{\epsilon_{O}}$=8.70} \\ 
		 \small{} & \small{}& \small{$\Delta\epsilon_{O}(\sigma)$=0.20, $\Delta\epsilon_{O}(\xi)$=0.13} \\ 		  	
		\hline
		\hline\\[-1.5ex]
					
 \end{tabular}
\end{table}

  \begin{table}[p!]
	\centering
	\caption{Detailed results of the silicon and oxygen abundance analysis line-by-line for the star BD+40 4210 ($T_{\rm eff}$=18300 K, log $g$=2.2 and $\xi(O)$=$\xi(Si)$=13.3 km~s$^{-1}$). }
	\label{target8}
		\begin{tabular}{lcc}
		\hline   
		\hline\\[-1.8ex]  
        \small{Line} & \small{EW (m$\AA$)} & \small{$\epsilon_{Si}\pm \Delta\epsilon_{Si}$ (dex)} \\    	
   		 \cline{1-3}\\[-1.5ex]
		\small{\ion{Si}{II} 6347} & \small{89$\pm$20} & \small{7.52$\pm$0.14} \\    
		\small{\ion{Si}{III} 4552} & \small{397$\pm$16} & \small{7.50$\pm$0.08} \\    	
		\small{\ion{Si}{III} 4567} & \small{340$\pm$20} & \small{7.51$\pm$0.16} \\         
		\small{\ion{Si}{III} 4574} & \small{212$\pm$15} & \small{7.46$\pm$0.14} \\  
		\small{\ion{Si}{III} 5739} & \small{370$\pm$20} & \small{7.53$\pm$0.01} \\
		 \cline{3-3}\\[-1.5ex]
   		\small{} & \small{} & \small{$\bar{\epsilon_{Si}}$=7.53} \\
   		\small{} & \small{} & \small{$\Delta\epsilon_{Si}(\sigma)$=0.01, $\Delta\epsilon_{Si}(\xi)$=0.09} \\   			    	
         \cline{1-3}\\[-2.0ex] 
         \cline{1-3}\\[-1.8ex] 	
         \small{Line} & \small{EW (m$\AA$)} & \small{$\epsilon(O)\pm \Delta\epsilon_{O}$ (dex)}\\    	
   		 \cline{1-3}\\[-1.5ex] 
		\small{\ion{O}{II} 4317} & \small{111$\pm$15}& \small{8.43$\pm$0.20} \\
		\small{\ion{O}{II} 4366} & \small{120$\pm$15}& \small{8.43$\pm$0.19} \\
		\small{\ion{O}{II} 4414} & \small{146$\pm$20}& \small{8.37$\pm$0.20} \\
		\small{\ion{O}{II} 4416} & \small{124$\pm$20}& \small{8.47$\pm$0.21} \\
		\small{\ion{O}{II} 4438} & \small{133$\pm$15}& \small{8.66$\pm$0.20} \\
		\small{\ion{O}{II} 4650} & \small{251$\pm$15}& \small{8.63$\pm$0.15} \\
		\small{\ion{O}{II} 4661} & \small{134$\pm$15}& \small{8.55$\pm$0.20} \\
		\small{\ion{O}{II} 4676} & \small{100$\pm$15}& \small{8.37$\pm$0.22} \\
		\small{\ion{O}{II} 6721} & \small{47$\pm$15}& \small{8.30$\pm$0.22} \\
		\small{\ion{O}{II} 4189} & \small{53$\pm$15}& \small{8.47$\pm$0.22} \\
		 \cline{3-3}\\[-1.5ex] 
		\small{} & \small{}& \small{$\bar{\epsilon_{O}}$=8.48}\\   
		\small{} & \small{}& \small{$\Delta\epsilon_{O}(\sigma)$=0.06, $\Delta\epsilon_{O}(\xi)$=0.07}\\  		 	
		\hline
		\hline\\[-1.5ex]	
			
		\end{tabular}
\end{table}

\section{Fitting line models}\label{app2}
\begin{figure*}[p!] 
   \centering
   \caption{Best-fit HHeSiO model ($T_{\rm eff}$=33000 K, log $g$=4.0 dex, $\xi(Si)$=$\xi(O)$=3.0 km~s$^{-1}$, log(Si/H)= -4.20 dex, log(O/H)= -3.30 dex) to the observed spectrum of J20314965+4128265. Si and O lines used in the analysis are indicated with dark and light blue vertical lines, respectively. H and He lines are indicated with dotted and solid black vertical lines for reference.}  
  \includegraphics[width=0.98\textwidth ]{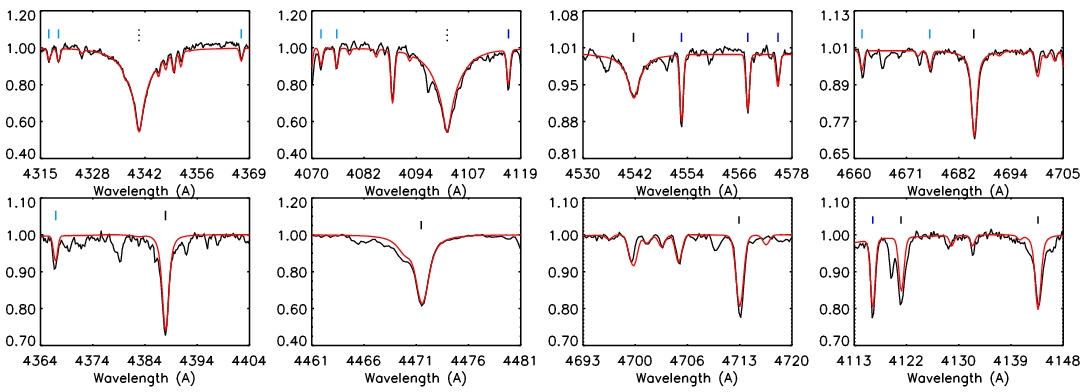}  
 \label{sp_models1}
\end{figure*}

\begin{figure*}[p!] 
   \centering
   \caption{Best-fit HHeSiO model ($T_{\rm eff}$=32000 K, log $g$=3.6 dex, $\xi(Si)$=$\xi(O)$=7.5 km~s$^{-1}$, log(Si/H)= -4.50 dex, log(O/H)= -3.30 dex) to the observed spectrum of J20295701+4109538. Sort lines mark the positions of Si, O, H and He lines, as in Fig~\ref{sp_models1}. }  
  \includegraphics[width=0.98\textwidth ]{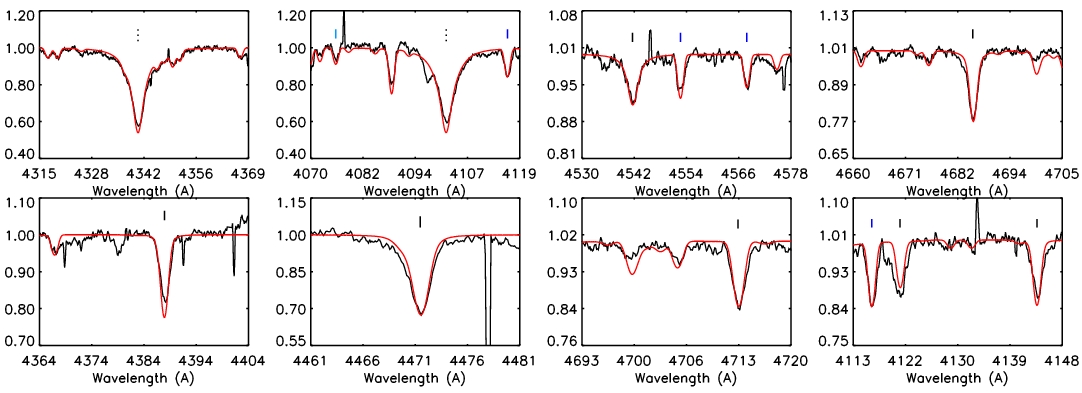}  
 \label{sp_models2}
\end{figure*}

\begin{figure*}[p!] 
   \centering
   \caption{Best-fit HHeSiO model ($T_{\rm eff}$=30000 K, log $g$=3.9 dex, $\xi(Si)$=$\xi(O)$=5.2 km~s$^{-1}$, log(Si/H)= -4.50 dex, log(O/H)= -3.30 dex) to the observed spectrum of J20272428+4115458. Sort lines mark the positions of Si, O, H and He lines, as in Fig~\ref{sp_models1}. }  
  \includegraphics[width=0.98\textwidth ]{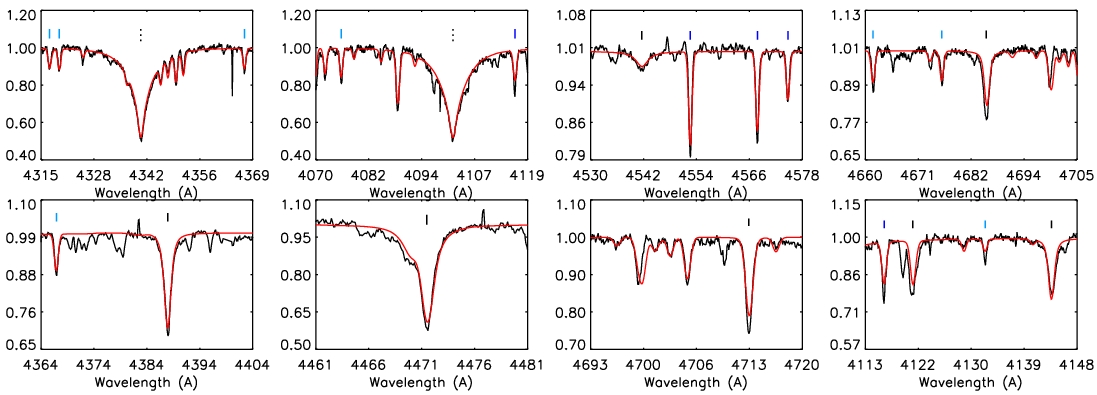}  
 \label{sp_models3}
\end{figure*}

\begin{figure*}[p!] 
   \centering
   \caption{Best-fit HHeSiO model ($T_{\rm eff}$=30000 K, log $g$=3.2 dex, $\xi(Si)$=$\xi(O)$=17.0 km~s$^{-1}$, log(Si/H)= -4.50 dex, log(O/H)= -3.40 dex) to the observed spectrum of J20334610+4133010. Sort lines mark the positions of Si, O, H and He lines, as in Fig~\ref{sp_models1}.}  
  \includegraphics[width=0.98\textwidth ]{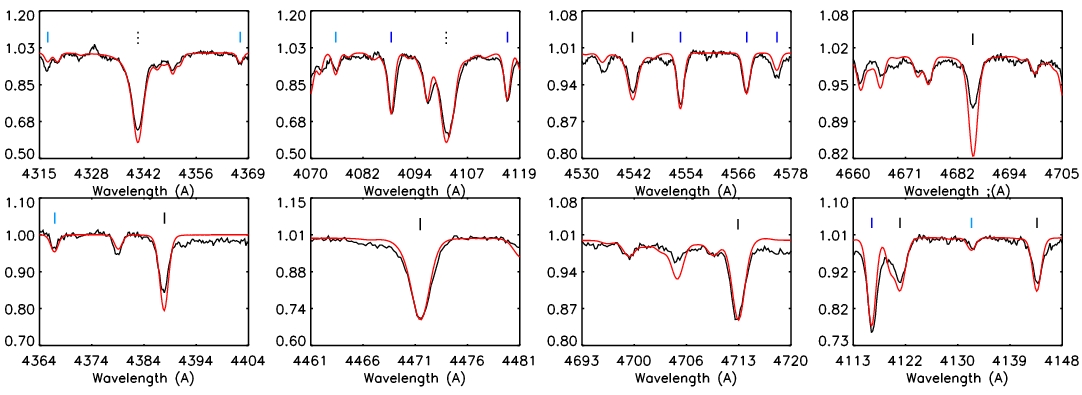}  
 \label{sp_models4}
\end{figure*}

\begin{figure*}[p!] 
   \centering
   \caption{Best-fit HHeSiO model ($T_{\rm eff}$=30000 K, log $g$=4.0 dex, $\xi(Si)$=$\xi(O)$=5.6 km~s$^{-1}$, log(Si/H)= -4.50 dex, log(O/H)= -3.30 dex) to the observed spectrum of J20314605+4043246. Sort lines mark the positions of Si, O, H and He lines, as in Fig~\ref{sp_models1}. }  
  \includegraphics[width=0.98\textwidth ]{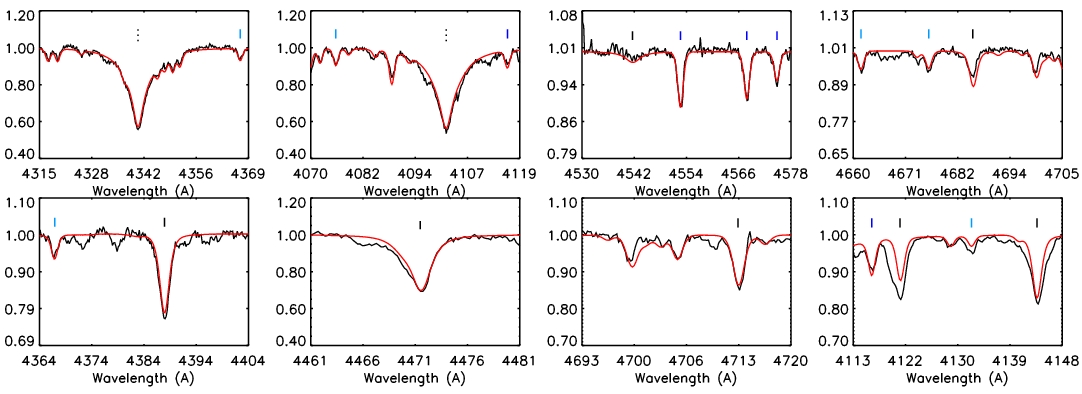}  
 \label{sp_models5}
\end{figure*}

\begin{figure*}[p!] 
   \centering
   \caption{Best-fit HHeSiO model ($T_{\rm eff}$=29000 K, log $g$=3.7 dex, $\xi(Si)$=$\xi(O)$=6.8 km~s$^{-1}$, log(Si/H)= -4.50 dex, log(O/H)= -3.30 dex) to the observed spectrum of J20292449+4052599. Sort lines mark the positions of Si, O, H and He lines, as in Fig~\ref{sp_models1}. }  
  \includegraphics[width=0.98\textwidth ]{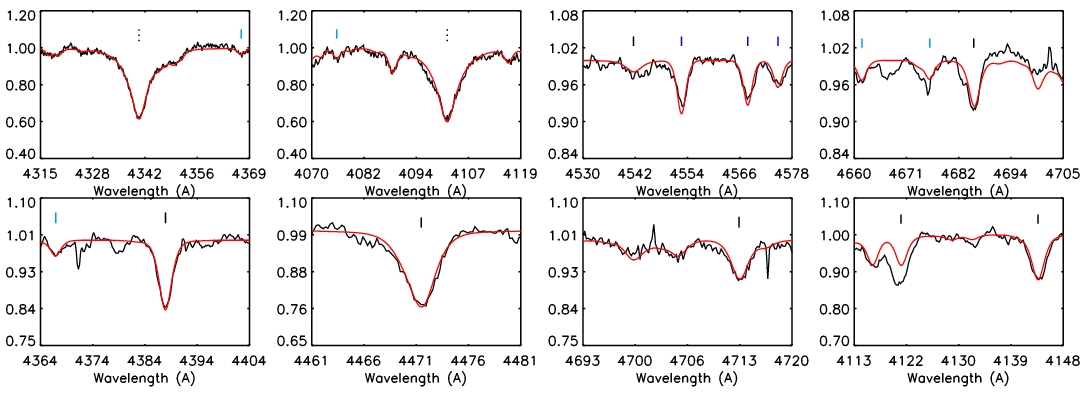}  
 \label{sp_models6}
\end{figure*}

\begin{figure*}[p!] 
   \centering
   \caption{Best-fit HHeSiO model ($T_{\rm eff}$=19000 K, log $g$=3.8 dex, $\xi(Si)$=$\xi(O)$=5.9 km~s$^{-1}$, log(Si/H)= -4.50 dex, log(O/H)= -3.30 dex) to the observed spectrum of BD+40 4193. Sort lines mark the positions of Si, O, H and He lines, as in Fig~\ref{sp_models1}.}  
  \includegraphics[width=0.98\textwidth ]{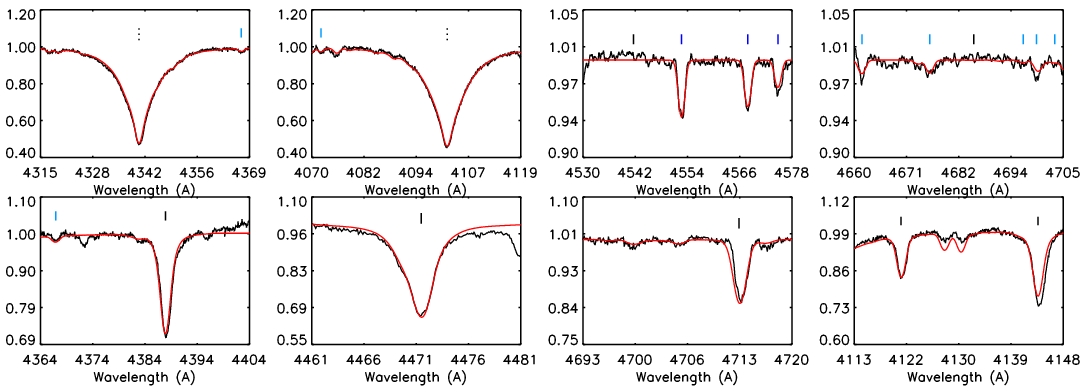}  
 \label{sp_models7}
\end{figure*}

\begin{figure*}[p!] 
   \centering
   \caption{Best-fit HHeSiO model ($T_{\rm eff}$=18300 K, log $g$=2.20 dex, $\xi(Si)$=$\xi(O)$=13.3 km~s$^{-1}$, log(Si/H)= -4.50 dex, log(O/H)= -3.40 dex) to the observed spectrum of BD+40 4210. Sort lines mark the positions of Si, O, H and He lines, as in Fig~\ref{sp_models1}. }  
  \includegraphics[width=0.98\textwidth ]{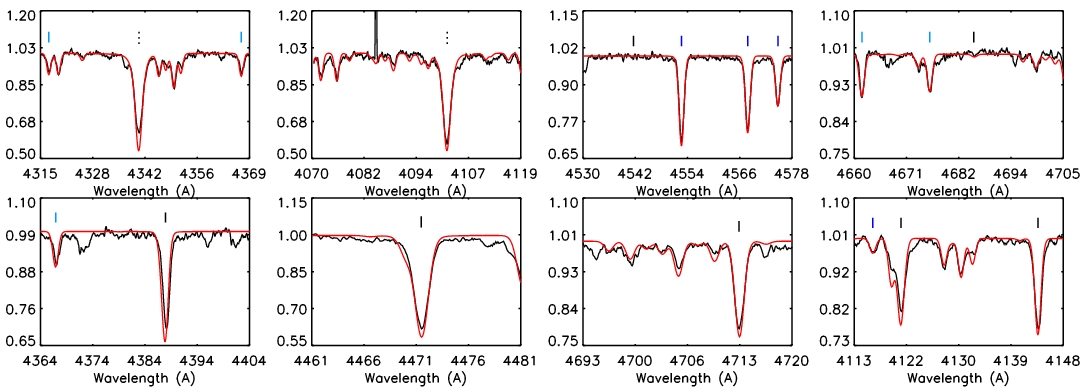}  
 \label{sp_models8}
\end{figure*}

Figures~\ref{sp_models1}-\ref{sp_models8} show, for each selected sample star, several H, He, Si and O spectral lines and the best fitting model derived from our abundance analysis. 

\end{appendix}

\end{document}